%% file: main.tex
\documentclass[conference]{IEEEtran}
\usepackage{url}
\usepackage{tikz}
\usepackage{amsmath}
\usepackage{booktabs}
\usepackage{multirow}
\usepackage{filecontents}
\usepackage{subfigure}
\usepackage{caption}

\begin{document}

\date{}

\title{ Seamless Website Fingerprinting in Multiple Environments }

\author{\IEEEauthorblockN{Chuxu Song}
\IEEEauthorblockA{Rutgers University\\
cs1346@scarletmail.rutgers.edu}
\and
\IEEEauthorblockN{Zining Fan}
\IEEEauthorblockA{Rutgers University\\
zf140@scarletmail.rutgers.edu}
\and
\IEEEauthorblockN{Hao Wang}
\IEEEauthorblockA{Rutgers University\\
hw488@cs.rutgers.edu}
\and
\IEEEauthorblockN{Richard Martin}
\IEEEauthorblockA{Rutgers University\\
rmartin@scarletmail.rutgers.edu}}


 

\maketitle

\begin{abstract}

Website fingerprinting (WF) attacks identify the websites visited over
anonymized connections by analyzing patterns in network traffic
flows, such as packet sizes, directions, or interval times using
a machine learning classifier. Previous
studies showed WF attacks achieve high classification accuracy.
However, several issues call in to question if existing WF approaches 
are realizable in practice and thus motivate a re-exploration.
Due to Tor's performance issues and resulting poor browsing experience
the vast majority of users opt for Virtual Private Networking (VPN)
in spite of VPNs weaker privacy protections. 
Many other past assumptions are increasingly unrealistic
as web technology advances. Our work addresses several key
limitations of prior art. First, we introduce a new
approach that classifies entire web-sites rather than individual web
pages. Site-level classification uses traffic from all
site components, including advertisements, multimedia,
and single page applications. Second, our Convolutional
Neural Network (CNN) uses only the jitter and size of 500
contiguous packets from any point in a TCP stream,
in contrast to prior work requiring heuristics to find page boundaries. 
Our seamless approach makes eavesdropper attack models realistic.
Using traces from a controlled browser, we show our CNN
matches observed traffic to a web-site with over 90\% accuracy.
We found the training traffic quality is critical as classification accuracy
is significantly reduced when the training
data lacks variability in network location, performance, and clients'
computational capability. We enhanced the base CNN's efficacy using 
domain adaptation, allowing it to discount irrelevant features,
such as network location. Lastly, we evaluate several defensive strategies
against seamless WF attacks.

\end{abstract}

\newcommand{\university}{University}
\newcommand{\vultr}{Cloud1}
\newcommand{\linode}{Cloud2}
\newcommand{\ocean}{Cloud3}
\newcommand{\optimum}{Cable}

\input{text/100_Introduction}
\input{text/125_Related_work}

\input{text/130_threat_model}

\input{text/150Methodology}

\input{text/600_evaluation}

\input{text/700_protection}

\input{text/900_Conclusion_Future_Work}

\label{sec:figs}



\bibliographystyle{IEEEtranS}
\bibliography{main}
\end{document}

%% file: text/100_Introduction.tex
\section{Introduction}
\label{sect:intro}


Identification of website destinations has been an important function for several decades. It has positive aspects in protecting users from malicious websites,
as network administrators can redirect or block such traffic. Website traffic identification \cite{DBLP:conf/cikm/KanT05,sahoo2019malicious,Kaur_2014,DraperGil2016CharacterizationOE} also plays a role in protective environments, such as primary schools,
and plays a role in digital marketing and analytics, network management, and regulatory compliance. Website traffic monitoring also has negative
uses, such as censorship and surveillance of journalists and human rights organizations. 

Our work revisits the field of Website identification called {\em Website Fingerprinting} (WF).
WF associates network flows with web pages or websites using pattern matching on traffic features visible
even when the data are encrypted and the headers obscured by intermediate routing approaches such as virtual private networks (VPN)s or 
The onion routers. The remaining visible features can include the packet size, timing, and direction, which can be learned or
recognized by a variety of machine learning and statistical techniques. 
We develop a deep learning model, dubbed WFNet, which expands state of the art in WF, showing how Convolutional Neural Networks (CNNs) \cite{NIPS2012_c399862d,oshea2015introduction} can be used to identify a destination website using just packet size, jitter, or both. We did not consider direction as the bi-bidirectionally of a flow in a router
is not as widespread as is believed. One study found that 65\% of route pairs have some asymmetry~\cite{he2005routing}. 

\label{sec:Threat Model} 
\begin{figure*}[t]
\begin{center}
\subfigure[Traditional WF Attack Threat Model]{\includegraphics[width=0.49\linewidth,height=5cm]{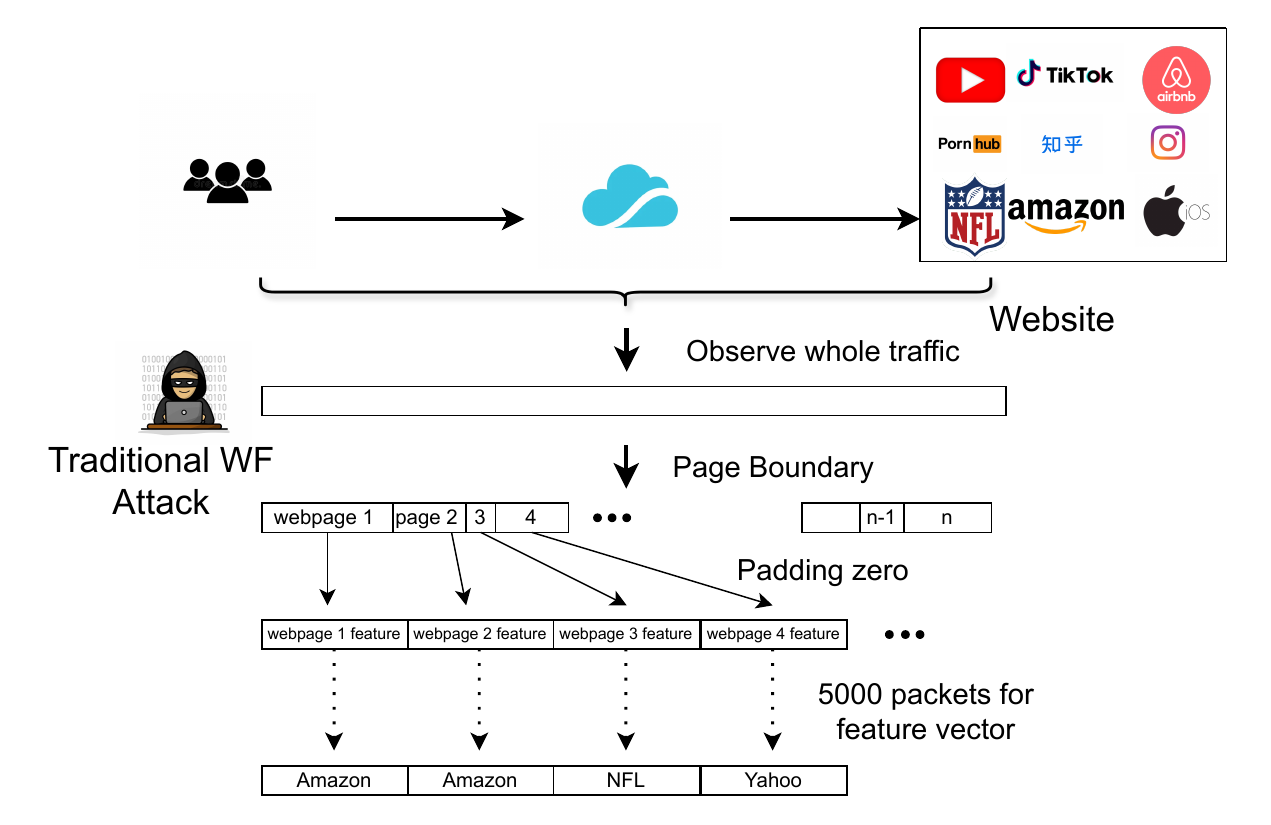}}
\hfill
\subfigure[Seamless WF Attack Threat Model]{\includegraphics[width=0.49\linewidth, height=4cm]{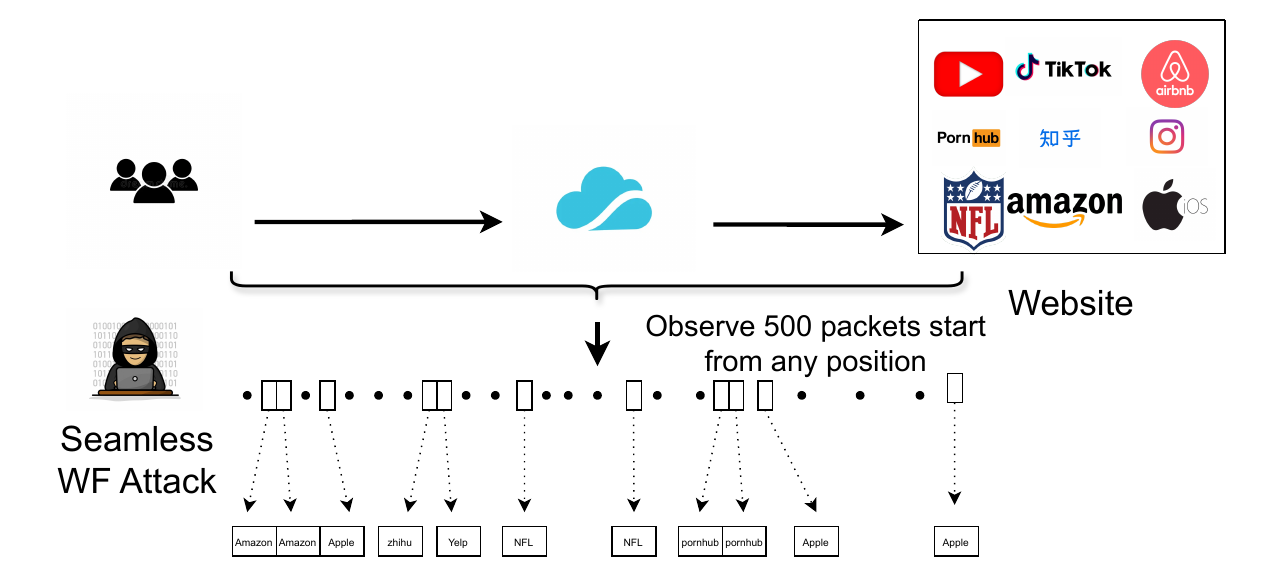}}
\caption{\label{fig:new threat model} }
\end{center}
\vspace{-0.6cm}
\end{figure*}

WF fingerprinting has a long history. In the beginning, 
protocols such as HTTPs were investigated by works such as ~\cite{liberatore2006inferring,bissias2006privacy,lu2010website}.
These showed it is possible to obtain accuracies of 60-90\% using statistical techniques such as Bayesian inference.
However, two factors motivate a renewal in examining WF and defenses to WF attacks. The first is that the expansion of the
Internet means that the current networking environment is more complex and variable than it was in the early 2010s.
Web sites moved from static HTML web pages to include complex sites with
per-user dynamic layouts, location-specific advertising, and large audio and video streams. 
Not accounting for these complexities can significantly
reduce the classification accuracy of prior approaches. Our results show that ignoring the environment's differences in the training phase 
can reduce accuracy by 30-50\%. Our work thus investigates if more modern techniques, such as large CNNs, can close the accuracy gap. 

After researchers showed that WF is effective when applied to standard protocols, attention moved to examining WF on the Tor network~\cite{dingledine2004tor} in such works as ~\cite{277132,10.1145/1655008.1655013,10.1145/2517840.2517851,wang2014effective}, and extensively documented in a recent
~survey\cite{aminuddin2023rise}. 
However, despite nearly 20 years of Tor's pivotal role in enhancing browsing privacy, its adoption has been limited. One recent study showed only two million active users per day worldwide~\cite{chen2019measuring}, out of 100's of millions of possible users. Most users opt for standard protocols such as HTTPS, VPNs, DNS over TLS, and DNS over HTTPS.
Tor's performance penalties and resulting low adoption are a second reason to revisit techniques that enhance standard protocols' privacy in the face of WF attacks in modern settings.

More specifically, our work addresses four limitations of prior approaches. The first two limitations arise from characterizing
WF is identifying specific web pages, rather than the entire domain of a website. 
Our WFNet uses the entire site as a class, thus including 
traffic from advertisements, images, sub-pages, and single-page applications, as opposed to some works that 
select individual page URLs or even classify sites by a single homepage~\cite{Rimmer_2018}.
A second related concern occurs when classifying web pages, which is that the 
page size becomes a defining feature. However, in practice, a passivity-observing attacker would have to split a TCP
stream at page boundaries, which is difficult in practice on protocols that keep the stream open between requests.
We have results that up to 30\% accuracy can be obtained by web page size alone as a defining feature, which
implies this feature is a significant contribution to overall accuracy. 

\textbf{Lack of experiment environments}. Many prior works are the representativeness of the training and testing sets,
which often uses a single client at one location. 
Our work shows that different network environments and locations in the network
have a significant impact on WF accuracy, as these have different bandwidths and latencies to a website.
We also show that the client's computing environment, such as the processor speed and available memory,
also has significant impacts on traffic patterns, and thus WF accuracy. We thus collected traffic from different network environments, including inside cloud data centers, within a University, and from home-based networks connected via cable modems.  

\textbf{Unrealistic attack model}. Many works assume the attacker views the entire
web transaction. Under these assumptions, a WF attack can view the beginning and end of a web page accesses 
or a set of accesses. We have a more relaxed attack model where the attacker only needs to observe 500 contiguous packets
anywhere over the entire session. For example, an attacker could simply record 500 packets in
a router without regard to the timing of the recording. Our approach is thus {\em seamless} in the sense that it does not require finding any boundaries in the traffic --- any contiguous segment can be used.

\begin{table*}[t]
\setlength{\tabcolsep}{1.9pt}
\caption{Networking and Execution environments for each client type.}
\label{table111}
\centering
\resizebox{\linewidth}{!}{  
\begin{tabular}{@{}llllllll@{}}
\toprule
                                    & \multicolumn{2}{l}{Network Maximums (Mbit/s)}            & \multicolumn{2}{l}{CPU info}  & \multicolumn{3}{l}{Geekbench}                                   \\ \midrule
                                    & Download speed & Upload speed & number of cores & memory (GB) & Single-Core Score & Multi-Core Score & HTML5 render (pages/sec) \\
\university          & 900                     & 880                   & 64              & 503         & 1384              & 12720            & 27.7                     \\
\vultr-Chicago       & 6700                    & 6000                  & 1               & 3.82        & 1797              & 1787             & 40.5                     \\
\vultr-Miami         & 5000                    & 4670                  & 1               & 3.82        & 1797              & 1785             & 40                       \\
\vultr-London        & 7840                    & 2300                  & 1               & 3.82        & 1793              & 1793             & 38                       \\
\linode-Toronto      & 2000                    & 1345                  & 2               & 3.83        & 746               & 970              & 18.8                     \\
\ocean-San Francisco & 4175                    & 2315                  & 2               & 3.83        & 731               & 1479             & 23.1                     \\
\optimum home 1      & 387                     & 26                    & 4               & 15.5        & 1297              & 4272             & 30                       \\
\optimum home 2      & 276                     & 36                    & 4               & 7.74        & 294               & 875              & 10.6                     \\ \bottomrule
\end{tabular}}

\end{table*}

Our experiments used synthetic traffic in a closed-world approach, which is when we assume the user's traffic is one of the sites in the training set.
We show that even within the constraints of a closed world evaluation, more realistic and unrestricted browsing environments significantly increase the difficulty of attacks.
To compare the observed accuracy of our WFNet with prior art, we tested the three mainstream WF attack models—CUMUL~\cite{Panchenko2016WebsiteFA}, DLWF~\cite{Rimmer_2018}, Triplet Fingerprint~\cite{277132}
on our dataset and found the accuracy rates of these models were all below 60\%. In contrast, by increasing the scale of our WFNet, we were able to achieve accuracy rates exceeding 90\%. 
An important contribution of our work is a demonstration that CNNs used for WF are powerful, stable generalizers when supported by extensive data.

We also show the key to obtaining high accuracy requires having a comprehensive training set that is representative of
many different network environments. For example, we used three distinct environments when 
running a headless browser to generate traffic: virtual machines in cloud providers,
a large server at a University, and desktops at home networks connected with cable modems. Similar to other machine learning results that require diverse training
data, having multiple network environments is critical to getting good classification accuracy\cite{sun2017revisiting,hestness2017deep}. In addition to the network environment, we also found that the
computational performance of the machine running the browser also has a high impact on accuracy. In particular, the home networking desktops were much older
machines with comparatively weak CPUs, and this generated traffic that was substantially different compared to cloud traffic. 
Training the CNN on traffic from the cloud providers and testing on home networks thus resulted in a low accuracy (around 50\%).

We found that domain adaptation can maintain accuracy when training traffic is acquired at different network points from the training and testing sets.
Domain adaptation builds a feature model on the interior of the CNN to model similar features across domains, in our case the traffic collection location.
The approach modifies the network training phase by adding an auxiliary network to the CNN which classifies the locations rather than the website
Intuitively, simultaneously maximizing the website accuracy while minimizing the location accuracy during training helps retain features related to website classification
while eliminating features related to the locations; such learning of location-invariant features would improve the model's generalization across different locations.

We also explored two privacy-enhancing techniques using evasion techniques. Evasion refers to altering the input to a trained neural network
to result in an incorrect classification. Here, the adversary is the entity with the trained network while the user tries to evade
correct classification. Our first approach, which we call {\em Inflation}, adds randomization delay in the jitter and
inflates the exiting packet sizes by randomized amounts. We found the delays and sizes needed to reduce accuracy using Inflation to be untenable for existing applications, however. 
A second approach, which we call {\em Active Injection}, adds extra packets at random. This approach resulted in large drops in accuracy, with 3\% additional
total bytes lowering the accuracy to 10\%. However, active injection requires more protocol changes than inflation. In particular, while inflation can
be performed straightforwardly in middleboxes without knowledge of the higher-level protocol, active injection requires either end-point changes or split-style TCP
connections.

\textbf{Organization.} 
The remainder of the paper is organized as follows. Section~\ref{sec:related} describes related work. Next, Section~\ref{sec:methodology} describes our methodology,
including our network environments, the amount of traffic collected, our vanilla WFNet, and a WFNet variant equipped with domain adaptation techniques. 
Section~\ref{sec:experiments}
shows our experiment results, including training on the full and partial traffic traces, traces taken across
aged traffic, and the impact of training set size on the number of websites and the amount of training traffic.
Section~\ref{sec:protection} describes two methods for obscuring the destination website by modifying packet timings or adding additional dummy packets.
In Section~\ref{sec:conclusion} we summarize our conclusions and speculate on what it would take to make networking protocols privacy robust in the face of machine learning
techniques.

\begin{table*}[t]
\caption{Network Characteristics of the different locations.}
\label{table112}
\scalebox{1.1}{
\begin{tabular}{@{}lcccccccc@{}}
\toprule
                                   & \university & \multicolumn{3}{c}{Cloud 1} & Cloud 2       & Cloud 3 & \multicolumn{2}{c}{Home cable} \\ \midrule
                                   &  NJ          & London  & Miami  & Chicago  & San Francisco & Toronto & location 1     & location 2    \\
Average inter-packet interval (ms) & 3959       & 9418    & 8708   & 10893    & 7185          & 6052    & 2604           & 4553          \\
Average packet size (Bytes)        & 2014       & 2670    & 2452   & 2711     & 2125          & 2579    & 1180           & 1281          \\
Average browser bandwidth(MB/s)    & 0.5        & 0.28    & 0.28   & 0.24     & 0.3           & 0.42    & 0.45           & 0.21          \\ \bottomrule
\end{tabular}}

\end{table*}

%% file: text/125_Related_work.tex
\section{Background and Related Work}
\label{sec:related}

\captionsetup{justification=justified,singlelinecheck=false}
\begin{table}[]
\caption{The table compares our dataset with traditional datasets, highlighting that ours includes data from multiple environments and a significantly higher number of unique URLs per website, ensuring enhanced diversity.}
\label{tab:different_datasets_comparison}
\scalebox{0.7}{
\begin{tabular}{@{}llllll@{}}
\toprule
                            & DF  \cite{sirinam2018deep}    & AWF  \cite{Rimmer_2018}          & Wang14\cite{wang2014effective}        &  WFIS \cite{Panchenko2016WebsiteFA} & Our work               \\ \midrule
Number of client environments & 1                  & 1              & 1                  & 1                                        & 8                      \\
Total training vectors              & 100000             & 2250000        & 18000              & 150000                                   & \textgreater{}6500000  \\
Vectors per website        & 1000               & 2500           & 100                & 2000                                     & \textgreater{}300000   \\
Total web pages              & 100000             & 2250000        & 18000              & 150000                                   & \textgreater 1 million \\
  Total unique pages        & N\textbackslash{}A & Home page only & N\textbackslash{}A & 150000                                   & 800000                 \\
Unique pages per website  & N\textbackslash{}A & Home page only & N\textbackslash{}A & \textless{}2000                          & 20000                  \\ \bottomrule
\end{tabular}
}
\end{table}

In this section, we first describe related work in general traffic classification. 
Next, we describe the limitations of existing website fingerprinting. We then
show related work in transfer learning and domain adaptation.


\begin{figure*}[t]
    \centering
    \includegraphics[width=0.98\textwidth]{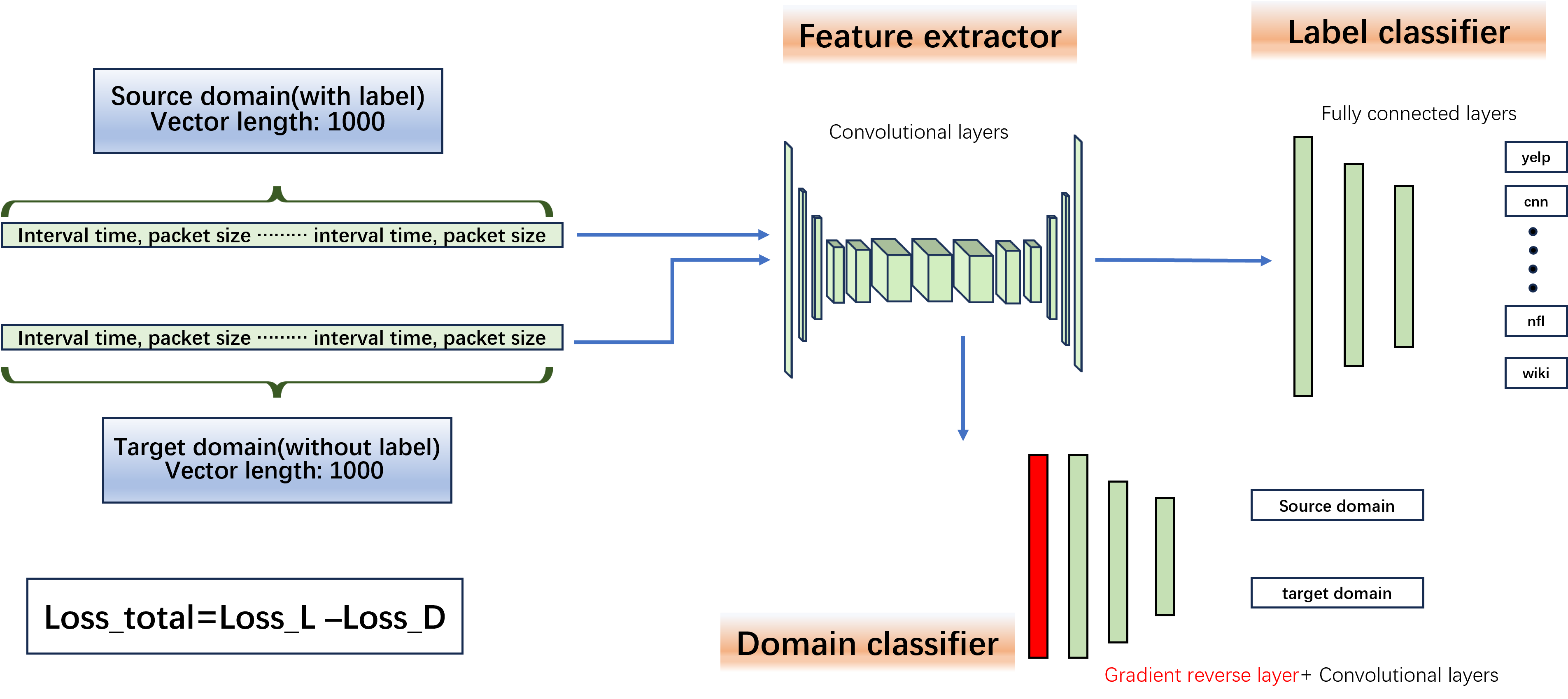}
    \caption{\label{fig:domainnetwork} Neural Network Structure Implementing Domain Adaptation.}
\end{figure*}

\subsection{Traditional Classifiers}
The common pattern of traditional network classifiers is that they
use machine learning or statistical classifiers to identify, separate, and label 
types of network traffic. There has been a wide range of foundational strategies,
from Support Vector Machines (SVNs), nearest neighbor clustering methods, and neural networks.
In the domain of computer networks, the typical feature space 
includes packet jitter, size, the values of header fields such as flags and ports, and the packet direction,
which can be towards or away from a given network.

Various studies \cite{lotfollahi2018deep,Rezaei_2020,7925139,10.1145/3447382} explored diverse approaches and feature sets. The Moore network classifier \cite{moore_network_traffic} employed a feature-rich set, achieving over 98\% classification accuracy for a range of traffic types. However, this approach faced complexity in feature selection and limited neural network structures to fully connected networks. The Lopez classifier \cite{lopez-2017-network} adopted CNNs and RNNs, achieving impressive accuracy but relying on network features that may not always be available. Williams' model \cite{williams2006preliminary} focused on packet parameters and achieved high accuracy, though the feature space was complex. Some research simplified features, as the model in~\cite{lim_packet_2019}, which used packet payloads and improved the accuracy of eight different applications, particularly using ResNet. Shahbaz \cite{google_rezaei_2018achieve} conducted a study using semi-supervised machine learning to classify Google services like Google Drive, YouTube, and Google Docs based on packet timing data. They achieved over 90\% accuracy for script-generated traffic and approximately 81\% for human-generated traffic, showcasing the potential of machine learning in web application classification. However, their dataset was limited to Google-based applications, not diverse websites. Lastly, Auld, Moore's work \cite{auld_2007_bayesian} utilized packet header content, including various features, in Bayesian neural networks, achieving a 68.8\% average accuracy for 10 traffic categories. These studies collectively demonstrate a rich landscape of approaches and trade-offs in network traffic classification. Burn's approach \cite{10001054} showcases the accuracy of CNN models in website classification, including classifying ten websites from a single location, which is a similar task as in this work. However, their accuracy was limited to 80\%. 

Specific to website fingerprinting, researchers have developed a myriad of approaches for conducting WF attacks, alongside corresponding defensive mechanisms ~\cite{197185,10.1145/2517840.2517851,10.1145/3457904}. Traditionally, these methods primarily rely on monitoring and analyzing the timing and size characteristics of network traffic. For instance, Herrmann et al.~\cite{10.1145/1655008.1655013}. introduced a statistical analysis-based approach that identifies target websites by constructing a statistical feature model of website visits. On the other hand, recent studies have begun to leverage machine learning technologies to enhance the accuracy and efficiency of attacks. The application of deep learning algorithms~\cite{sirinam2019triplet,Rimmer_2018}, for example, allows for the automatic extraction of features from encrypted network traffic and website identification, demonstrating the feasibility of website fingerprinting attacks even amidst encrypted traffic. A quintessential example of this evolution is the method proposed by Panchenko et al.~\cite{Panchenko2016WebsiteFA}, which utilizes Support Vector Machines (SVM) to achieve high recognition rates in complex network environments.

\begin{figure*}[t]
    \centering
    \includegraphics[width=\textwidth]{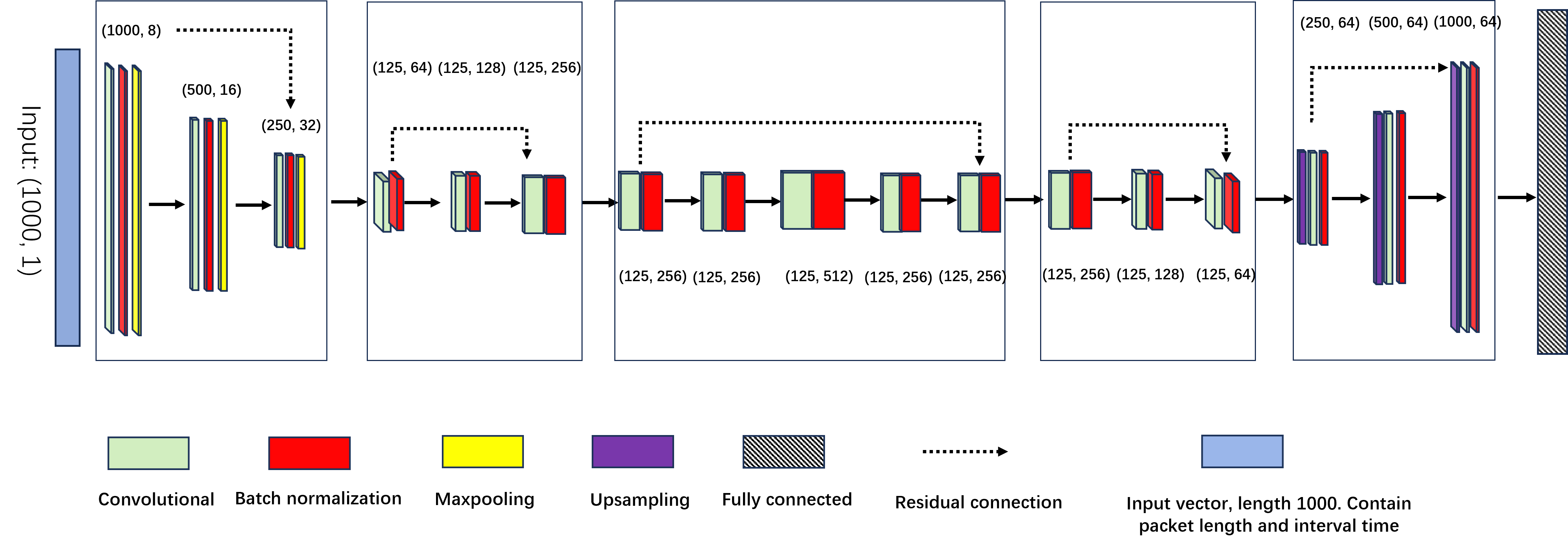}
    \caption{\label{fig:basicCNN} Structure of the WFNet-Base}
\end{figure*}

\begin{figure}[t]
    \centering
    \includegraphics[width=0.47\textwidth]{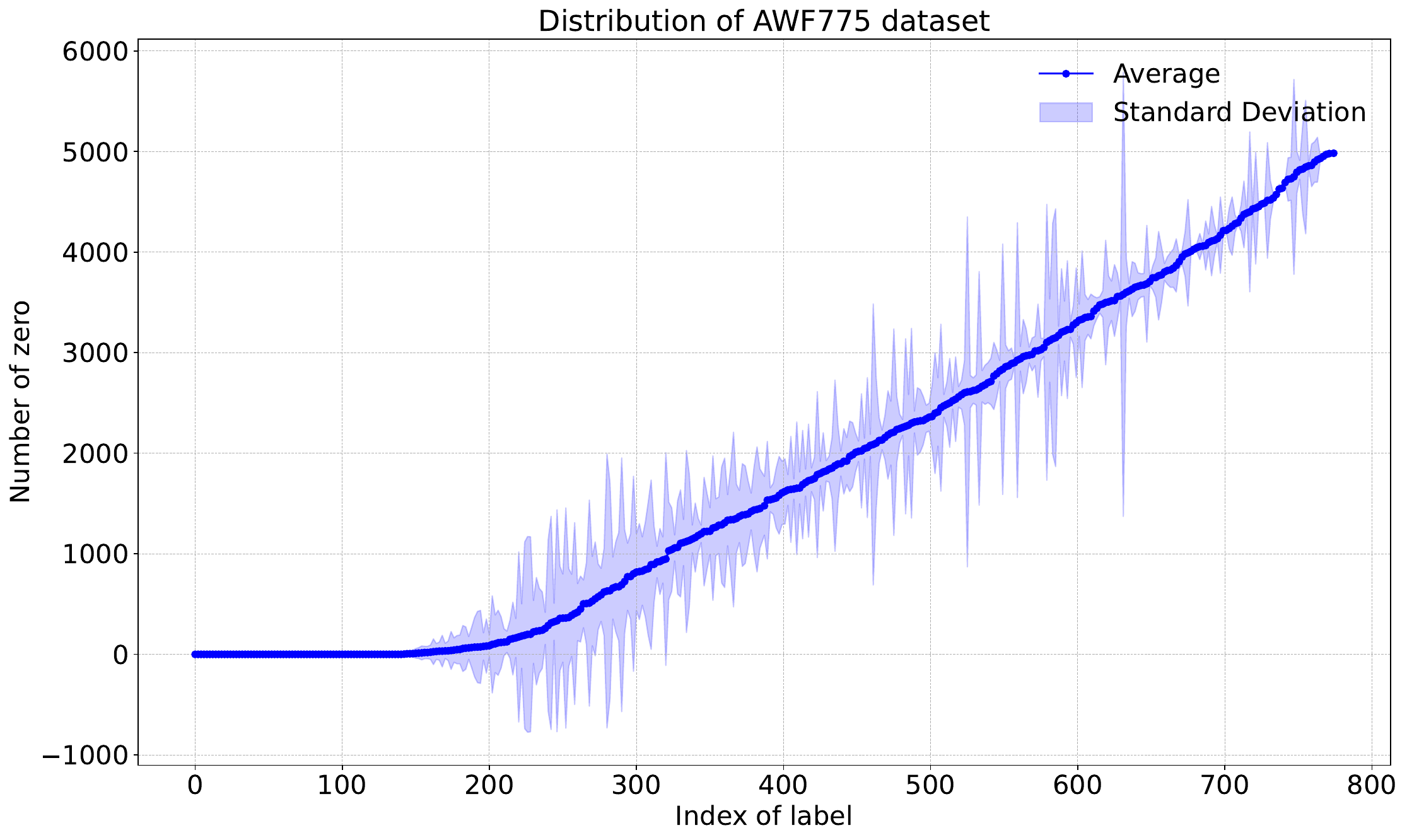}
    \caption{\label{fig:Wang_distribution} 
This figure illustrates the distribution of padding zeros across each class in the AWF775 dataset. Labels are ordered by the average padding zeros per instance, sorted in descending order based on the data volume generated by each website. The blue line indicates the mean number of padding zeros, while the shaded area represents the standard deviation.}
\end{figure}

\subsection{Machine Learning Techniques }

The field of machine learning has developed three techniques relevant to our work: Transfer learning, Domain Adaptation, and Backdoor learning. We give background on each of these techniques in turn.

\noindent
{\bf Transfer Learning}. Transfer Learning is a paradigm that leverages knowledge learned from one domain (source domain) to improve performance in another, often related, domain (target domain). Some notable approaches and contributions in this field include\cite{zhuang2020comprehensive}.
In our context, we use the ``pretrain-finetune" paradigm, which typically results in more robust models when presented with entirely untrained new datasets using zero-shot reasoning or after fine-tuning with limited labeled data.
This approach will enable us to effectively tackle diversity and complexity, thereby enhancing our WFNet's performance and generalization capabilities. 

\noindent
\textbf{Domain Adaptation}. Domain adaptation addresses the disparities in data distributions across different domains. The main objective of domain adaptation is to mitigate the adverse effects of
domain shifts on the model's performance. For example, a domain shift becomes visible when changing browsing from Chicago to Miami. Changing location may result in the model learning features of the location,
as opposed to the features of the website. 

When dealing with training data that encompasses multiple diverse domains, the incorporation of domain indices has proven helpful in mitigating differences among domains. This approach assists in balancing the gaps between various domains, thereby enabling better adaptation to the data of the target domain. Furthermore, certain models \cite{xu2023domainindexing,xu2023graphrelational,NEURIPS2021_45017f65} have the capability to smoothly transition between domains during continuous domain shifts~\cite{wang2020continuously,liu2023taxonomystructured}.

\begin{table*}[t]
\centering
\caption{\label{tab:locations} Accuracy of the WFNet Base trained on one location and tested on another.}
\scalebox{1.0}{
\begin{tabular}{lcccccccc}
\hline
               & \university         & \vultr-Miami         & \vultr-Chicago       & \vultr-London        & \linode              & \ocean       & \optimum home 1     & \optimum home 2      \\ \hline
\university         & {\textbf{0.91}} & 0.41                & 0.52                & 0.33                & 0.38                & 0.44                & 0.33               & 0.25                \\
\vultr-Miami    & 0.4                 & {\textbf{0.95}} & 0.87                & 0.56                & 0.6                 & 0.53                & 0.27               & 0.22                \\
\vultr-Chicago  & 0.56                & 0.84                & {\textbf{0.97}} & 0.5                 & 0.62                & 0.57                & 0.32               & 0.29                \\
\vultr-London   & 0.24                & 0.53                & 0.56                & {\textbf{0.95}} & 0.42                & 0.49                & 0.21               & 0.23                \\
\linode-Toronto         & 0.38                & 0.63                & 0.53                & 0.52                & {\textbf{0.97}} & 0.81                & 0.33               & 0.39                \\
\ocean-San Francisco  & 0.47                & 0.6                 & 0.65                & 0.5                 & 0.84                & {\textbf{0.96}} &      0.36              &           0.47          \\
\optimum home 1 & 0.38                & 0.35                & 0.31                & 0.28                & 0.44                & 0.49                & {\textbf{0.9}} & 0.66                \\
\optimum home 2 & 0.33                & 0.29                & 0.29                & 0.15                & 0.45                & 0.50                & 0.63               & {\textbf{0.89}} \\ \hline
\end{tabular}}
\end{table*}

\noindent
\textbf{Backdoor Learning}. Backdoor learning focuses on understanding and mitigating vulnerabilities associated with machine learning models that are manipulated to produce erroneous outputs when exposed to specific, crafted inputs. Prior research demonstrated that attackers can manipulate the output of machine learning models and reduce their accuracy by injecting a small fraction of malicious data into the training dataset~\cite{turner2019cleanlabel,li2022backdoor,Wang_2022_CVPR}. In our context,
clients can improve privacy by incorporating backdoor learning approaches that alter their traffic patterns, thus reducing the accuracy of trained models.

%% file: text/130_threat_model.tex
\section{Seamless WF vs Traditional WF}
In this section, we delve into the limitations of traditional website fingerprinting (WF) and examine how seamless WF effectively addresses these challenges. Concurrently, we introduce seamless WF's corresponding threat model. 

\subsection{Limitations of Traditional WF}
\noindent

WF has a long history, from it's origins As Tor is explicitly designed to provide anonymity and privacy, almost all of subsequent WF research focused on Tor traffic. The massive attention to Tor created a blind spot in the research community because the robustness of other protocols for fingerprinting attacks has been neglected.

A recent survey of WF attacks on Tor identified nine limitations of prior work: closed-world modeling, sequential browsing,
isolated traffic, replicability, traffic parsing, passive webpage, disabled cache, static content, and single webpage~\cite{aminuddin2023rise}. 
These assumptions limit the practicality of WF in real-world scenarios. Their work showed that most WF on Tor studies
applied these assumptions despite requiring controlled environments or laboratory experiments. We describe each of these limitations and how our work addresses them.

{\bf Closed vs Open World Modeling}
In the realm of website fingerprinting the distinction between the "closed world" assumption, where users only visit monitored sites, and the "open world" assumption, where users may access sites beyond the monitored scope, significantly impacts both the choice of classifier and the accuracy of classification. Previous research predominantly utilized homepages as representations of websites. Researchers assumed that in a closed-world setting, a classification task involving a thousand websites would yield high accuracy rates for Website Fingerprinting (WF) attacks and subsequently shifted their focus towards challenges in open-world scenarios. However, we contend that representing a website by a single or a few webpages does not adequately reflect the site's traffic patterns. Many websites now feature diverse content styles within their pages—such as Amazon, which incorporates both e-commerce and streaming video elements. 

To redefine the evaluation methodology for website fingerprint attacks, we propose a novel approach where, for each website, we visit most or all of its subpages to represent it comprehensively. Starting in a closed-world setting, we explore the feasibility of this approach. As depicted in Table~\ref{tab:different_datasets_comparison}, our dataset, on average, visits over 20,000 unique subpages per website, a number vastly surpassing the total number of unique webpages visited in other traditional datasets. Therefore, even though our dataset is designed for a closed-world analysis, it exhibits far greater diversity and provides a more realistic representation of website fingerprinting's effectiveness in real-world conditions.

{\bf Unrealistic Browsing Behaviors}
Modern browsers and technologies shifted from discrete page/document interactions to continuous user engagement. As discussed earlier in the context of Single Page Applications and multimedia content, various user actions, such as mouse movements, keyboard inputs, and scrolling, can significantly influence both the size and pattern of web traffic. For instance, mouse movements often trigger the loading of new content, resulting in additional data flow. This implies that using webpages as standalone instances may not effectively represent a website's traffic profile, as users engage in diverse actions across different subpages.

To better simulate realistic conditions, our approach to generating website data involves the use of Puppeteer~\cite{puppeteer} to simulate a wider range of user actions. This method allows for a more dynamic and representative model of user interaction, capturing the complexities of modern web usage more accurately. By enhancing the realism of our simulated traffic, we aim to improve the robustness and applicability of website fingerprinting methodologies.

{\bf Unrealistic Obserability}
In the majority of public datasets, researchers often process complete packet sequences into single-page packet sequences in a laboratory setting, aiming for the attack model to handle traffic from individual webpages. Notably, when webpages generate varying numbers of packets, these sequences are typically padded with zeros to uniform lengths, usually around 5000 packets~\cite{sirinam2019triplet,277132}, for classifier input. We have observed that such padding introduces additional distinguishing features that significantly aid classification. Previous studies have suggested that after deploying models in real-world settings, attackers could leverage features like packet interval time gaps to split packet sequences and isolate single-webpage data. However, the increasing popularity of single-page applications (SPAs) and the rise of streaming media sites have blurred the boundaries between webpages. Users might spend several hours on a single page, which substantially reduces the frequency and timeliness of attacks that require observation of complete web session traffic or its initial data. Furthermore, the emergence of streaming media platforms like TikTok and YouTube Shorts, where each swipe to a new video is akin to navigating to a new page, complicates this scenario. To ensure user experience, these sites often preload subsequent webpage videos, leading to a continuous data stream with minimal noticeable time gaps, thereby significantly increasing the potential noise associated with webpage splitting. Our approach sidesteps the boundary issue because the level of classification is higher, at the entire site, rather than needing to discriminate individual pages.

\textbf{Using Page Size as a Feature}:
In Figure~\ref{fig:Wang_distribution}, we present the distribution of padded zeros within a 5000-element vector that represents page size in the AWF775 dataset~\cite{Rimmer_2018}. A near-linear relationship is evident, significantly enhancing the accuracy of our predictive models. However, as previously discussed, the act of splitting traffic in modern web environments introduces significant challenges. We contend that padding zeros is unrealistic, thus  leading to a substantial overestimation of the efficacy of website fingerprinting in practical scenarios. Whilezero-padding facilitates model training and initial testing, its utility diminishes in real-world applications where dynamic content and adaptive traffic behaviors prevail. Thus, reliance on such methods inflate performance metrics, misrepresenting the true robustness of website fingerprinting techniques under typical network conditions.

\textbf{Overlooking Differences in Network and Computing Environments}:
We found that the network environment and computing power of the client greatly influence traffic features and most prior work ignores this variability. The most importance factors we found are: 

\begin{enumerate}
    \item Physical Location:
     Location impacts round trip latency, and thus timing and packet order, routing symmetry, as well as actual content. For example, differences arise in packet traffic due to targeted advertising that is based on location. 
    \item ISP providers:
      Even for users located in close physical proximity, for example, in neighboring datacenters, the choice of ISP can greatly impact network traffic features. We found different ISP supported different Maximum Transmission Units (MTUs), for example. ISPs also have different kinds of routers which impacts bandwidth, latency, and packet ordering. 
    \item Client Capability:
      First, computationally weak clients generate many different patterns compared to strong ones across all feature sets. One example we saw in practice arose from resources clients that supported
      TCP Segmentation Offload (TSO) capabilities. We observed notable differences in website loading rates and sequences compared to their non-TSO counterparts.
      
  \end{enumerate}
  Table~\ref{tab:different_datasets_comparison} compares our environments to prior works. The key differences are the scale of traffic collected per web site, the total traffic volume, and the number of client environments. All other works used a single environment, and concentrated more on the expanding the number of sites rather than collecting more traffic to improve per-site accuracy.   
  
  One work that addresses some of the above challenges was Cherubin\cite{277132}, which proposed an innovative approach to website fingerprinting within the Tor network, emphasizing the collection of data generated by different users at the network's exit relays. By leveraging information exposed through DNS queries, they could obtain labels for the websites currently being accessed by users, facilitating the creation of a representative and diverse dataset. This methodology allows for the real-time update of the attacker model using online learning techniques, adapting to new data as it becomes available. However, the emergence of encryption protocols like DNS over HTTPS (DoH) increasingly complicates attackers' ability to acquire user traffic labels, even at exit relays. Moreover, While Cherubin's work assumes the feasibility of constructing a robust and generalizable model from a handful or a dozen instances, it does not account for the potential vast variability in packet distribution across different environments. Nor does it consider the virtually infinite number of real-world conditions, which can significantly affect the model's accuracy and generalizability. 

\subsection{New Threat Model}
\label{sec:Threat Model}
In the conventional website fingerprinting threat model, researchers typically utilize data from a complete single webpage session as features. This presupposes that attackers can start monitoring the victim's data transmission from the moment the data generation begins and can accurately identify the boundaries of webpages within the data stream, subsequently padding all collected data to a predetermined length for classification purposes. However, suppose users spend extended periods on boundary-less single-page applications. In that case, the effectiveness of such attacks may be significantly reduced, and the attacker may not be able to capture the victim's web browsing activities promptly.

In light of these challenges, we propose a simplified website fingerprinting threat model that is more suited for the era of streaming media. This model assumes that classification can be effectively achieved with fewer packets (500 fixed-length packets, as opposed to the state-of-the-art 5,000 packets). In this threat model, attackers can initiate monitoring at any time without being hindered by additional noise and latency associated with identifying webpage boundaries. This allows the model to consider frequent user interactions, subpages, page advertisements, and other elements as representations of the website's traffic

%% file: text/150Methodology.tex
\section{Methodology}
\label{sec:methodology}

In this section, we describe our method of synthetic trace generation. Our approach is broadly similar to prior works as we use a browser to access a set of known websites.
We process the recorded packet traces into feature vectors, which are used to train a WFNet. We first describe the website corpus, browser control, and URL selection. We next 
document the different networking environments. We then describe the model structure, how we altered the baseline model to support domain adaptation, and our pretraining and finetuning approach.

\subsection{Website Corpus}
\setlength{\tabcolsep}{2.9pt}

\begin{table}[]
\caption{This figure illustrates the composition of our dataset, highlighting the 42 websites from which browsing data was collected in a university setting. The red color represents 20 websites for which we collected data in eight different environments.}
\label{tab:websites_corp} 
\begin{tabular}{@{}llllll@{}}
\toprule
websites                              & label                         & websites                             & label                            & websites      & label    \\ \midrule
{\color[HTML]{FE0000} yelp.com}       & {\color[HTML]{FE0000} yelp}   & {\color[HTML]{FE0000} pinterest.com} & {\color[HTML]{FE0000} pinterest} & whatsapp.com  & whatsapp \\
{\color[HTML]{FE0000} britannica.com} & {\color[HTML]{FE0000} brit}   & {\color[HTML]{FE0000} wikipedia.org} & {\color[HTML]{FE0000} wiki}      & uber.com      & uber     \\
{\color[HTML]{FE0000} cnn.com}        & {\color[HTML]{FE0000} cnn}    & {\color[HTML]{FE0000} zhihu.com}     & {\color[HTML]{FE0000} zhihu}     & instagram.com & ins      \\
{\color[HTML]{FE0000} foxnews.com}    & {\color[HTML]{FE0000} foxn}   & {\color[HTML]{FE0000} iq.com}        & {\color[HTML]{FE0000} iq}        & shopify.com   & shop     \\
{\color[HTML]{FE0000} nfl.com}        & {\color[HTML]{FE0000} NFL}    & {\color[HTML]{FE0000} bilibili.com}  & {\color[HTML]{FE0000} bili}      & ubereats.com  & ubereats \\
{\color[HTML]{FE0000} nytimes.com}    & {\color[HTML]{FE0000} NYT}    & {\color[HTML]{FE0000} fandom.com}    & {\color[HTML]{FE0000} fandom}    & foodal.com    & foodal   \\
{\color[HTML]{FE0000} reddit.com}     & {\color[HTML]{FE0000} reddit} & {\color[HTML]{FE0000} taboola.com}   & {\color[HTML]{FE0000} taboola}   & bing.com      & bing     \\
{\color[HTML]{FE0000} target.com}     & {\color[HTML]{FE0000} target} & {\color[HTML]{FE0000} amazon.com}    & {\color[HTML]{FE0000} amzn}      & google.com    & google   \\
{\color[HTML]{FE0000} yahoo.com}      & {\color[HTML]{FE0000} yahoo}  & {\color[HTML]{FE0000} pornhub.com}   & {\color[HTML]{FE0000} pornhub}   & adidas.com    & adidas   \\
{\color[HTML]{FE0000} apple.com}      & {\color[HTML]{FE0000} apple}  & {\color[HTML]{FE0000} xvideos.com}   & {\color[HTML]{FE0000} xvideos}   & nike.com      & nike     \\
kiehls.com                            & kiehls                        & weather.com                          & weather                          & etsy.com      & etsy     \\
indeed.com                            & indeed                        & bestbuy.com                          & bestbuy                          & tiktok.com    & tiktok   \\
ikea.com                              & ikea                          & expedia.com                          & expedia                          & youtube.com   & youtube  \\
bbc.com                               & bbc                           & airbnb.com                           & airbnb                           & facebook.com  & facebook \\ \bottomrule
\end{tabular}
\end{table}

In order to have a representative sample of website traffic, we selected sites based on popularity. Table~\ref{tab:websites_corp} displays the 40 websites included in our dataset. The twenty websites highlighted in red were collected from all 8 environments and were used for the experiments in Sections~\ref{sec:diff} and ~\ref{sec:da}. The majority of these websites were selected from those with the highest traffic in the United States as in~\cite{similarweb}. We also included a few websites that have a similar nature (e.g., Bili, similar to YouTube, is a video-sharing website from China).

\subsection{Traffic Generation and Collection}
For web scraping and page navigation, we employed Puppeteer, which is a library to remotely control a browser. Using a browser rather than specialized downloaders, e.g., {\em wget}, generates more
representative traces as it will have realistic page loading, JavaScript execution, and advertisement rendering. We did not have an accurate model of human think-time and select behaviors but
instead sought to get a comprehensive data set via randomization. We extracted all links within each website and filtered out URLs with the same domain, assembling them into a pool of web pages. We attributed different weights to the pages based on their occurrence frequency in the pool and employed random sampling to select the next page for navigation. This process was iteratively repeated until a sufficient amount of data was acquired.

We utilized the {\em tshark} program to filter out all packets related to HTTPS. We extracted the packet arrival time and the size of each packet, storing them as 32-bit floating-point numbers within an array. The machine learning model's input comprised arrays of length 1000, equivalent to 500 packets, as each array sample consists of a tuple of the inter-arrival time and packet size. For every website collected from each location, we accumulated two binary files each exceeding 80 MB in size, with each value occupying 8 bytes. This resulted in a total of 10 million packets, forming 20,000 input vectors. During training and testing within a single location, data from one file was employed as the training set, while the other served as the testing set.

\subsection{Networking and Execution Environments}
To get traffic from different networking environments we deployed eight servers in six locations across three countries, with client connections spanning
five different ISP networks. Table~\ref{table111} shows details of each server.
We first characterize the three environments, \university, Cloud, and \optimum. We further used three different cloud providers,
and different locations, with two locations outside the United States; one in Canada and one in the United Kingdom. We characterized the
environments by both their networking parameters as well as the strength of the processor running the headless browser.  We selected 25 popular websites that are 
diverse in their functionality, for example, shopping, news, and entertainment.

Tables~\ref{table111} and ~\ref{table112} show the diversity of both the networking and execution environments. Download speeds varied by an order of magnitude, from close to 8 Gbs to less than 267 Mbs, and upload speeds likewise spanned a large range. To characterize computational power we used Geekbench \cite{Geekbench}, which shows a wide range of computational performance. In particular, the \university is unique in that the browser ran on a very large server, which is not typical for human-generated traffic. The inexpensive cloud virtual machines have similar computational power as a more modern desktop, as shown by their Geekbench CPU scores as well as GeekBench's browsing pages/second score. The \optimum home 2 location was run on a very old desktop, which is why its scores are low. However, this weak client with limited connectivity has a significantly distinct pattern compared to the other locations. Table~\ref{table112} shows the measured values of the actual browser traffic. The table shows that the average jitter, packet sizes, and bandwidths of these different locations vary by a factor of 2 on real browser traffic, thus showing these environments show a substantial performance range.  

To collect the \university traffic we employed two Dell R7525 computers located in a campus datacenter as clients. Concurrently, we installed the latest version of Headless Chromium on these two clients to facilitate browser-level operations using Puppeteer \cite{puppeteer}. These servers are connected to the \university network via a high-speed switch, vlan2300, with approximately 900MB/s upload and download speeds. This machine is equipped with two AMD EPYC 7452 32-core Processors, boasting significantly higher multi-core Geekbench scores compared to others. Within this setup, we gathered PCAP files at one-minute intervals. Each monitoring port yielded a total of 60 files, with these files being overwritten by new data on an hourly basis. This data collection was achieved through monitoring these two ports.

Furthermore, for the cloud service providers \vultr, \linode, and \ocean, (names have been anonymized)  we independently rented dedicated CPUs situated in five cities across three countries to serve as clients. In the cloud services provided by \vultr, we rented servers from Miami, Chicago, and London for use as clients in browsing websites. In Miami and Chicago servers, we utilized the same CPU specifications, one core from the AMD EPYC-Milan Processor, along with 4GB of memory. However, in the London server, we adopted a dual-core configuration with 8GB of memory. On the other hand, \ocean's servers employed Intel Xeon 2nd Generation Scalable processors, situated in San Francisco, utilizing two cores. As depicted in Table~\ref{table111}, the actual performance of a single core of the Intel Xeon processor is roughly half that of the AMD EPYC-Milan architecture. In \linode 's server, we employed two cores from the AMD EPYC 7501 processors located in Toronto, Canada, as clients. The single-core performance of this server is similar to that of the \ocean server. All three cloud service providers offered exceptionally high network bandwidth, exceeding two gigabytes per second in both upload and download speeds.

At the \optimum home 1 location, we utilized a Hewlett-Packard PC equipped with an Intel i7 4770 processor, and at \optimum home 2 location the desktop had an Intel Core2 Quad Processor Q6600 connection. Both servers had similar network bandwidth capabilities, boasting approximately 300MB download speed and 30MB upload speed. However, it's worth noting that the CPU performance of the i7 4770 significantly outperformed that of the Q6600, with the i7 4770 achieving a Geekbench score that was ten times higher than that of the Q6600.

\subsection{Model Structure}
\label{sec:modelstructure} 
We built a CNN-based model, referred to as WFNet-Base, to include 17 convolutional layers connected to three fully connected layers, as shown in Figure~\ref{fig:basicCNN}. We introduced batch normalization after each convolutional layer, aimed at normalizing the distribution of features within the network to enhance its adaptability to different input data. Furthermore, we organized the convolutional layers into five residual blocks. Within each residual block, we employed a residual connection strategy, connecting the first and last layers. This design allows us to increase the depth of the model while effectively improving the network's stability and convergence, resulting in superior performance when dealing with complex data. As a result of these modifications, the model's accuracy was significantly enhanced, achieving an accuracy of 93\% for website classification with the same two million samples. Additionally, we expanded upon the WFNet-base by adding six convolutional layers and increasing the number of channels in each layer, resulting in WFNet-Large with over 20 million parameters. WFNet-Base achieved higher accuracy in the mix location task(more training samples) mentioned in section 4.3.

\subsection{Domain Adaptation}
\begin{figure*}[t]
\begin{center}
\subfigure[Chicago]{\includegraphics[width=0.24\linewidth]{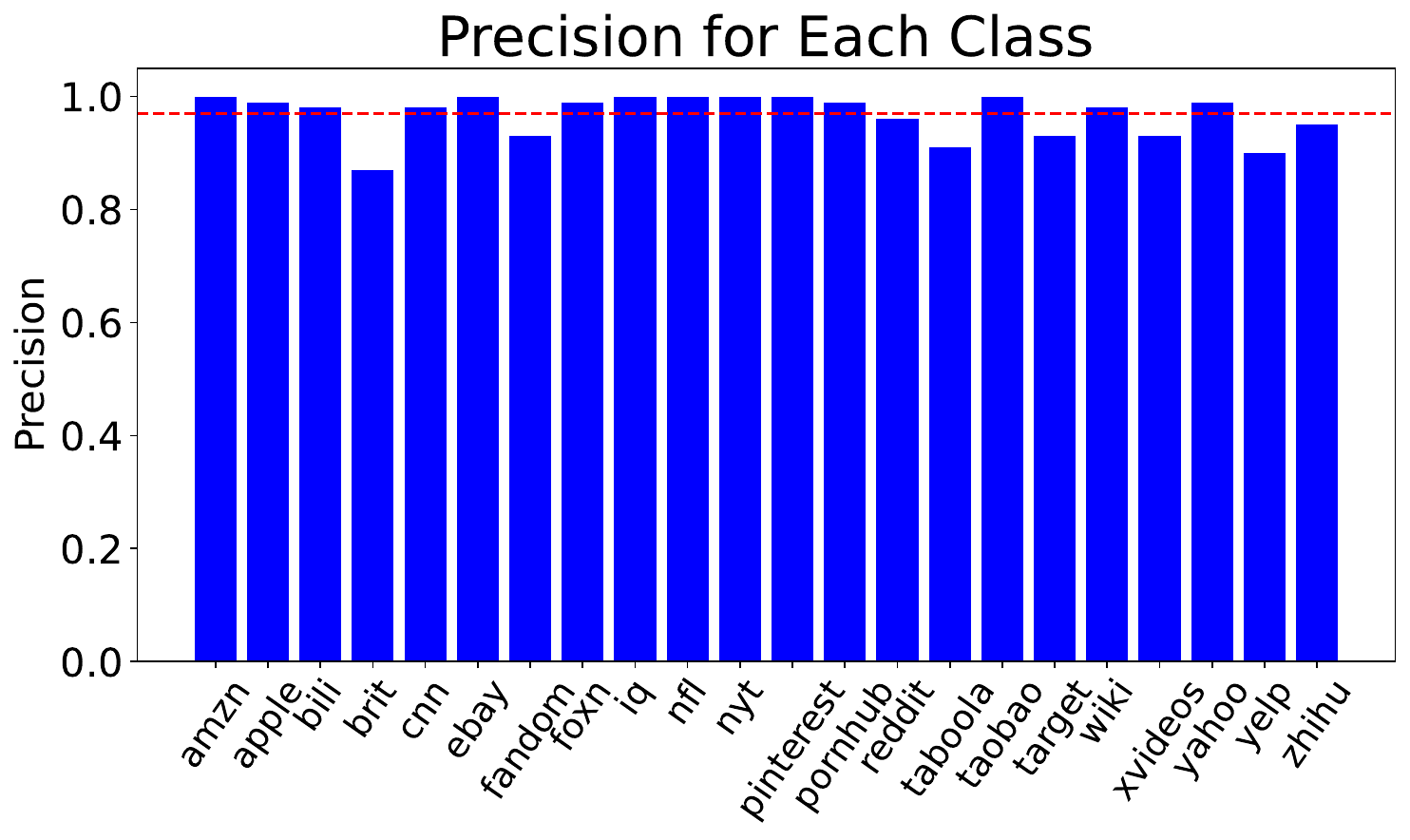}}
\subfigure[Miami]{\includegraphics[width=0.24\linewidth]{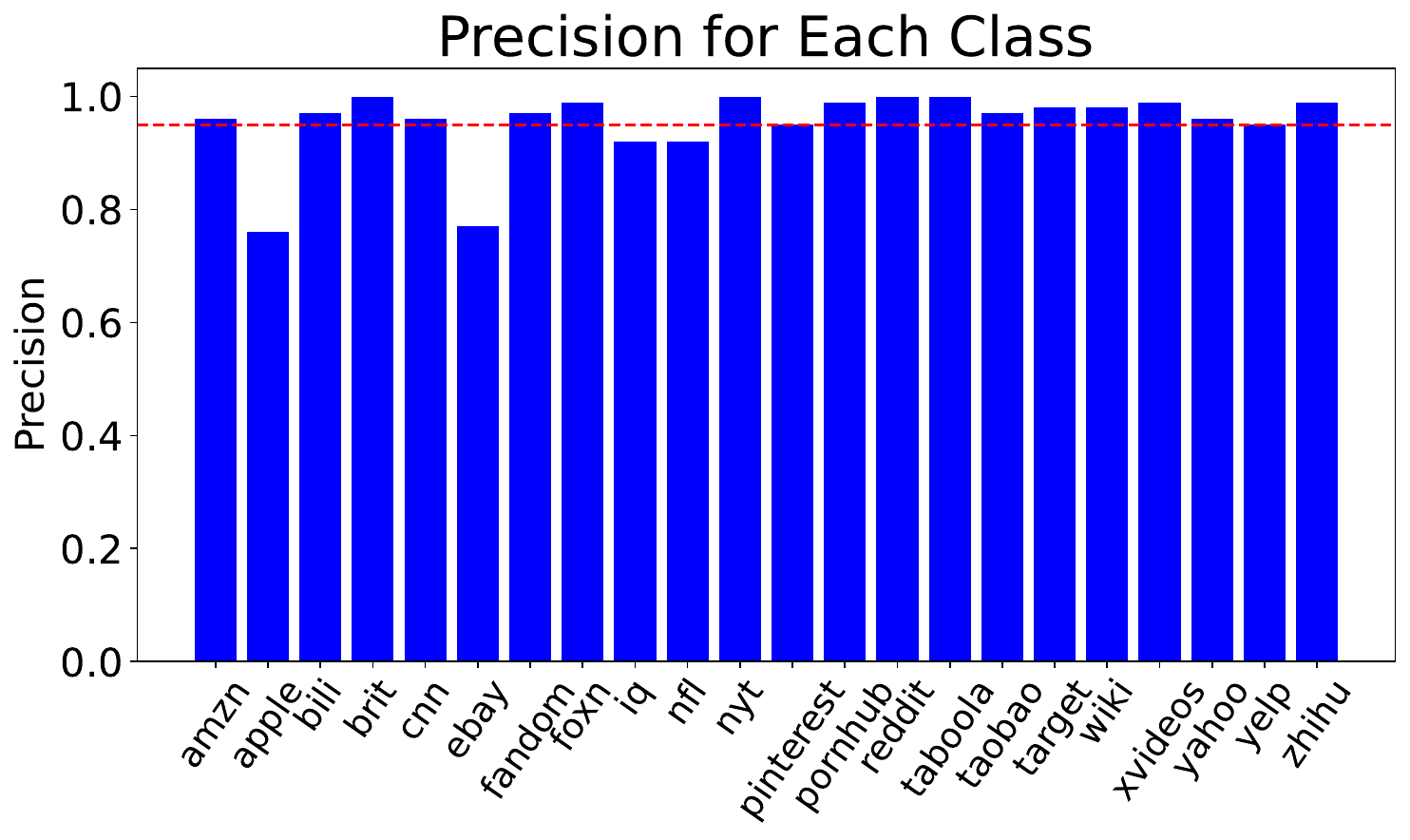}}
\subfigure[London]{\includegraphics[width=0.24\linewidth]{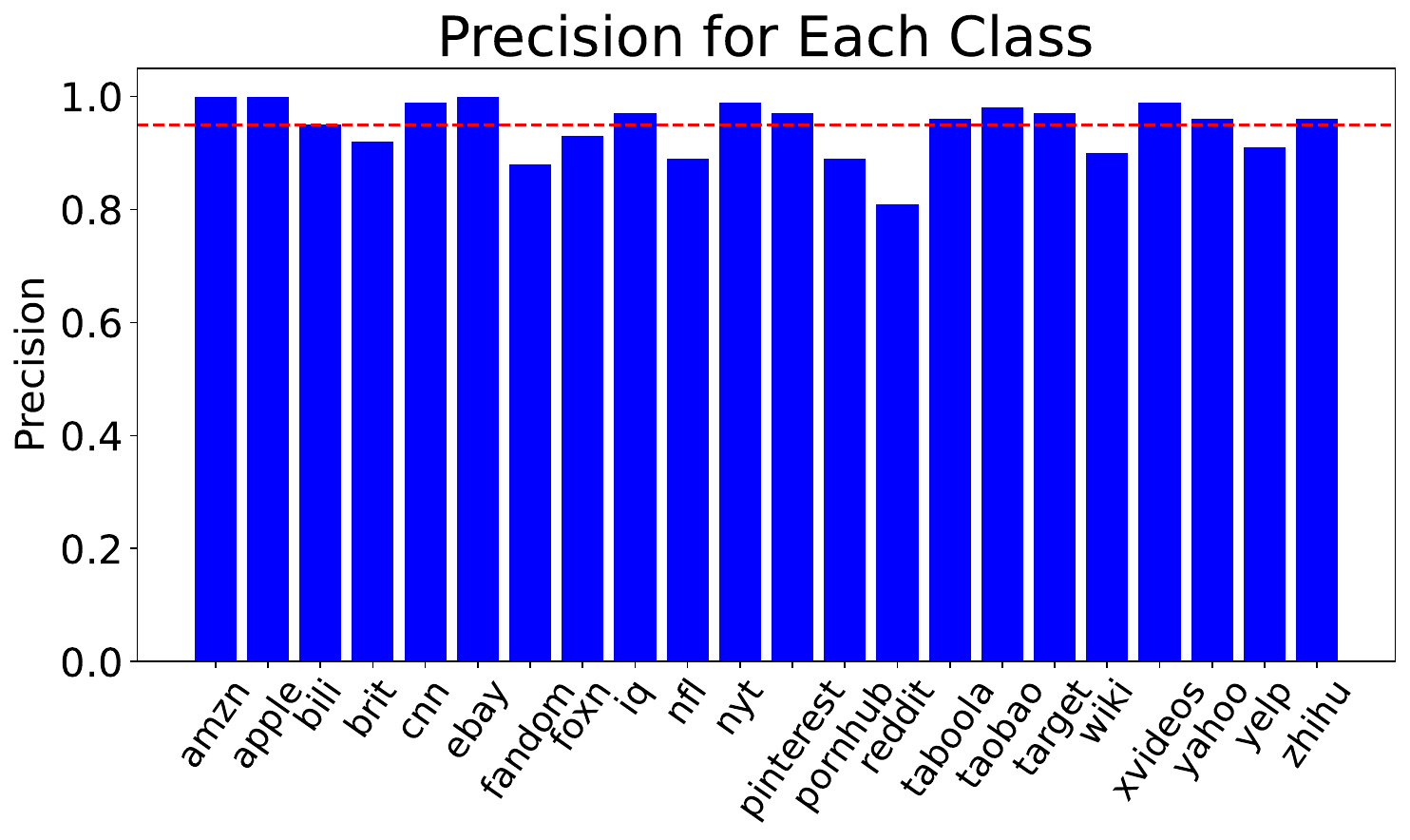}}
\subfigure[University]{\includegraphics[width=0.24\linewidth]{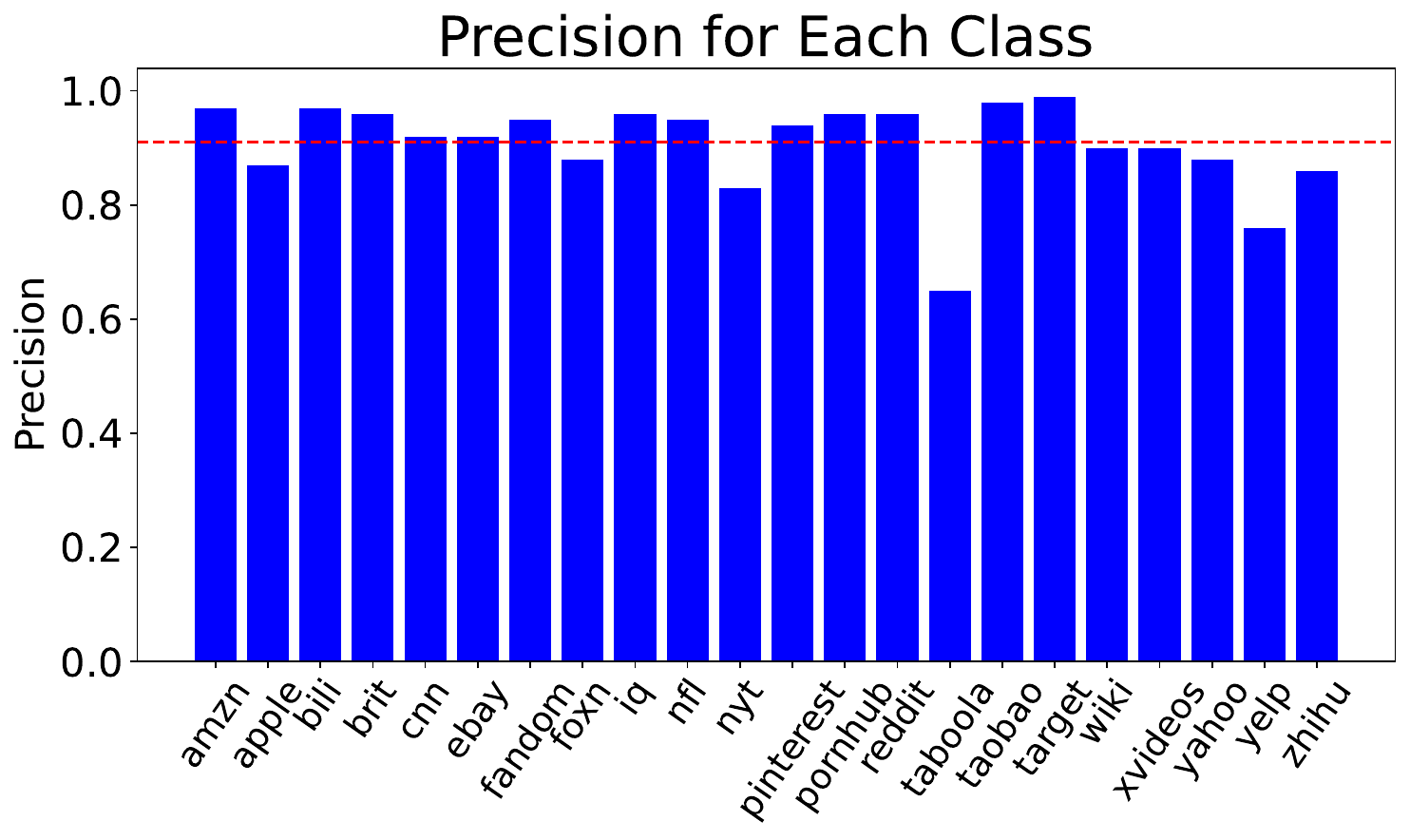}}
\subfigure[Cloud 3]{\includegraphics[width=0.24\linewidth]{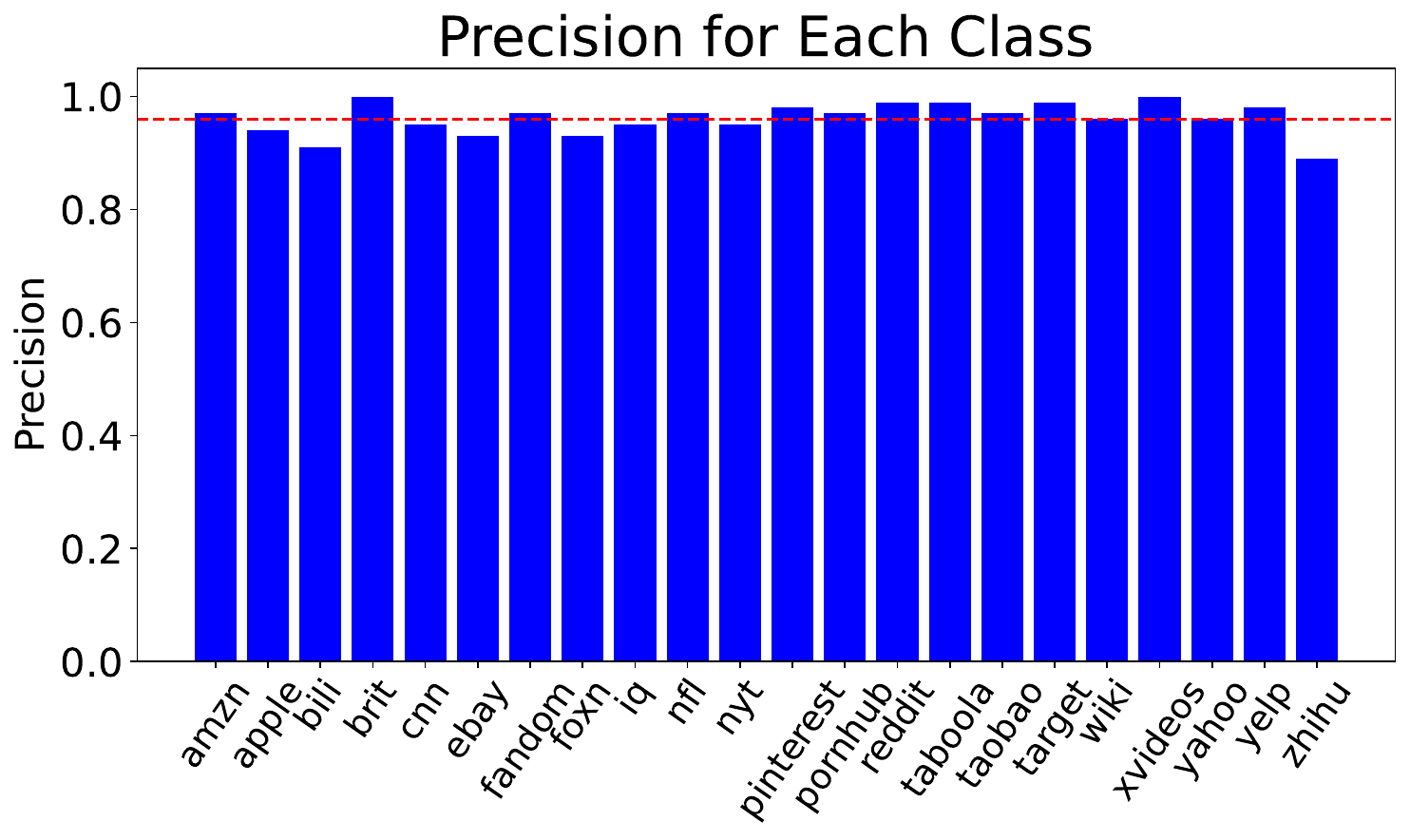}}
\subfigure[Cloud 2]{\includegraphics[width=0.24\linewidth]{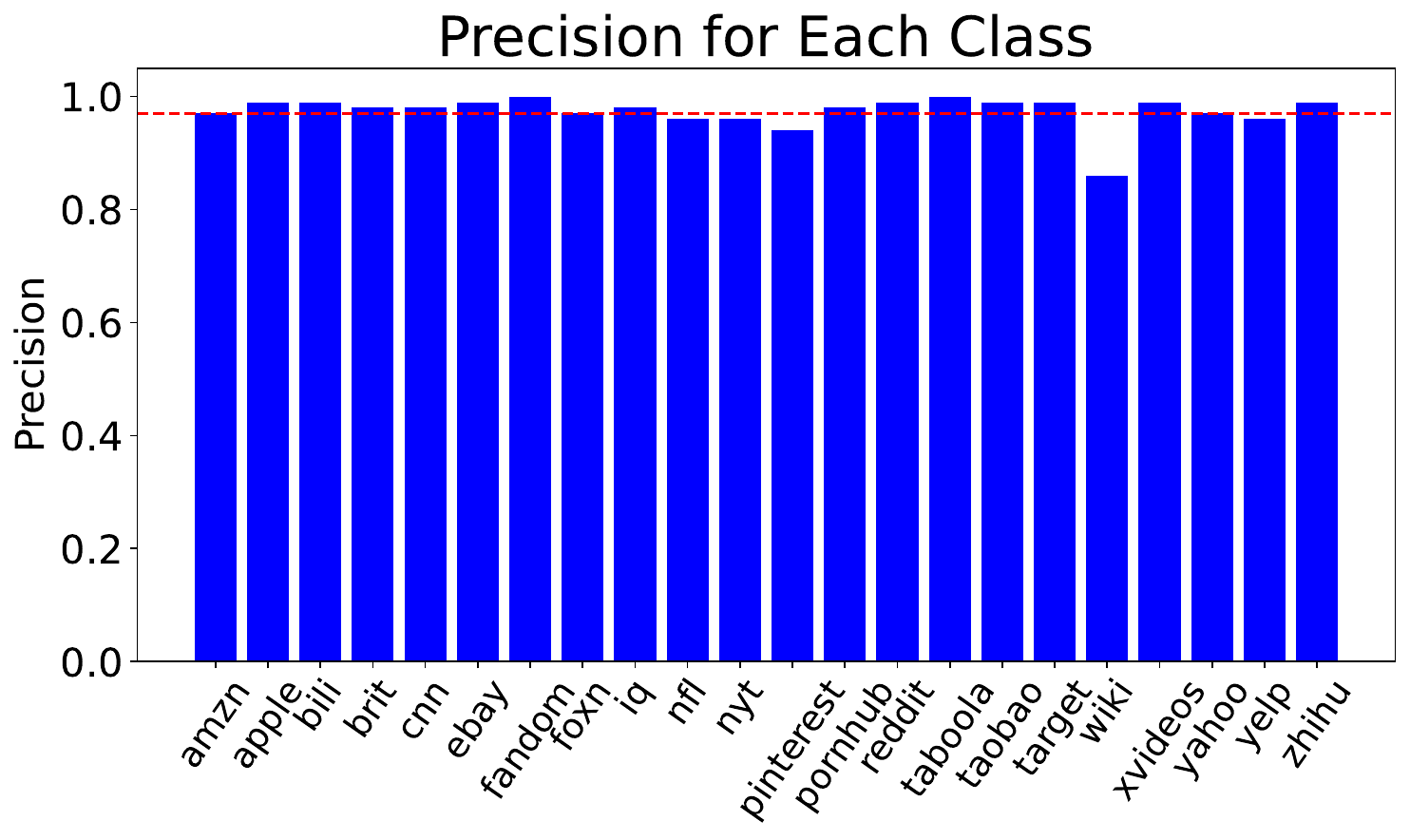}}
\subfigure[Home cable 1]{\includegraphics[width=0.24\linewidth]{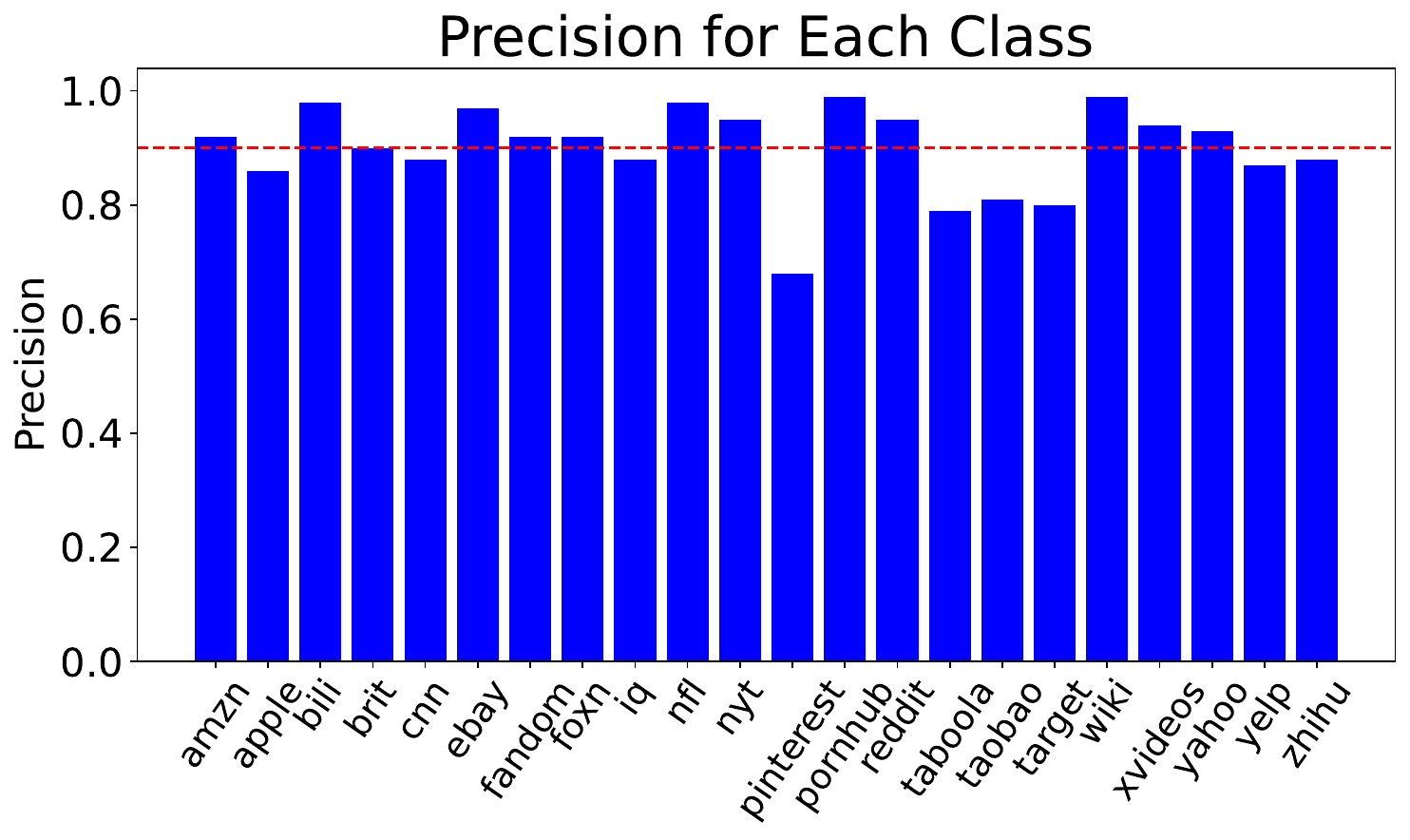}}
\subfigure[Home cable 2]{\includegraphics[width=0.24\linewidth]{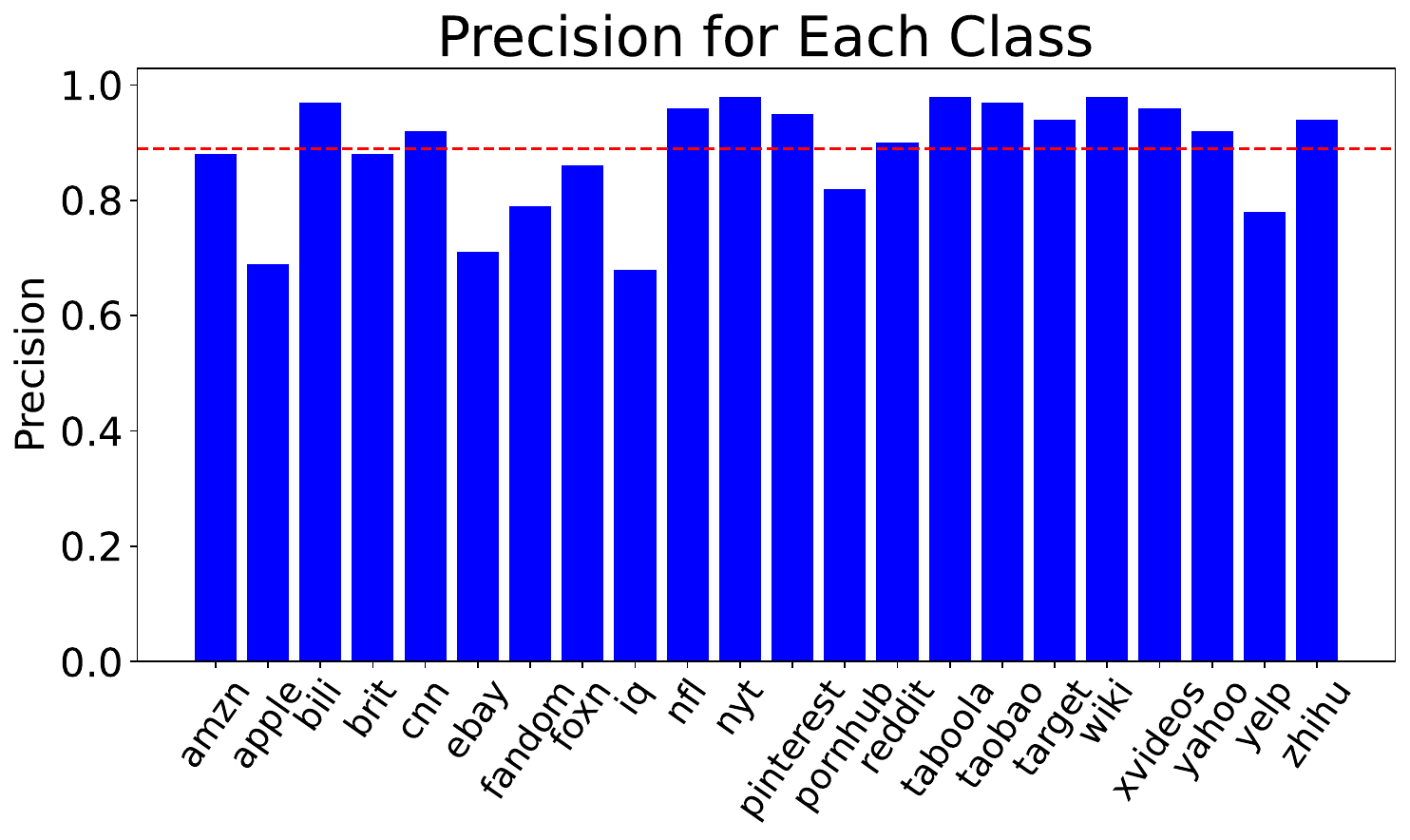}}
\caption{\label{fig:single-location}Accuracy of training on one location and testing on that location. Each location has 22 websites.}
\end{center}
\vspace{-0.3cm}
\end{figure*}

\begin{table}[]
\caption{Accuracy for various mainstream WF attack models within our closed-world seamless WF environment. The models include 'WFNet Base' and 'WFNet Large' which are high-parameter deep learning models developed by our team. } 
\label{tab:comparison}
\scalebox{0.85}{
\begin{tabular}{@{}llllll@{}}
\toprule
                 & CUMUL     & DLWF & Triplet Fingerprint & WFNet Large  & WFNet Base    \\ \midrule
10 website task  & 0.13      & 0.25 & 0.75                & 0.92        & 0.94        \\
20 website task  & 0.1       & 0.2  & 0.55                & 0.92        & 0.92        \\
Number of params & N/A (SVM) & 100K & 1 millions + KNN    & 23 millions & 10 millions \\ \bottomrule

\end{tabular}
}

\end{table}

A deep learning model trained on data from one ISP location usually achieves low accuracy (around 45\%) when tested on data from another ISP location. This is expected, as a training corpus needs to be large and diverse
for models to learn the wide variety of traffic patterns. 
However, taking into consideration that websites may have fixed ad banners and dynamic code, these features could lead one to believe that packet flows across different locations, network speeds, and time instances differed mostly in their temporal arrivals. Approaching the problem as a time-adjustment issue is flawed, however, as different computation speeds, latency and bandwidths result in different TCP packet sequences from its congestion control algorithm. Thus, instead of trying to time-adjust packets from different locations, we explore domain adaptation as a strategy for the network to learn the alignments of data streams across various scenarios and account for different network and client performance. 
When we possess labeled data streams generated from various locations and train our model accordingly, the model can predict data streams for all scenarios with exceptional accuracy, exceeding 95\%.

As in Figure \ref{fig:domainnetwork}, we consider data collected from different cloud servers as distinct domains, and our objective is to extract some domain-invariant features between the source domain $x_s$ (e.g. Chicago) and the target domain $x_t$, (e.g., Miami) such that a classifier trained on the source domain can work well in the target domain. We utilized the initial 20 convolutional layers with residual connections of our WFNet-Base model as a feature extractor $F$ and mapping packet feature $X$ to latent encoding $e= F(X)$. Then, we employ two separate fully connected networks as website classifiers $W$ and domain classifiers $D$. The website classifier takes encoding $e_s=F(x_s)$ and uses it to predict the website category. On the other hand, the domain classifier takes encoding $e_t$ and $e_s$ from both the source and target domains and tries to distinguish between them. Additionally, we introduce a gradient reverse layer at the beginning of the domain classifier, which multiplies the current gradient by a negative number during backpropagation. Overall, we want to min-max the following loss function:
\begin{equation}
    \min _{W, F} \max _D L_f(W, F)-\lambda_d L_d(D, F),
\end{equation}
where $L_f(W, F)$ denotes loss of feature extractor, $L_d(D, F)$ denotes the loss of the website classifier, and $\lambda_d$ is a hyperparameter that balances the two loss functions. 

Intuitively, we aim to minimize the label classifier loss while maximizing the domain classifier loss. In this way, the model will attempt to extract shared (domain-invariant) features between the target domain and the source domain. With these domain-invariant features, we expect that a classifier trained on the source domain to generalize well in the target domain. 
The website classifier is trained using source data with labeled website categories. Then, we assign different domain IDs (e.g., `0' means `source domain' and `1' means `target domain') to the source domain data and the target domain data, and train the domain classifier using such domain IDs (note that these are different from website category labels). Such domain adaptation training does not require any labeled target domain data; therefore an attack model trained on the source dataset can be easily reused on various servers and user terminals.

When integrating datasets from various scenarios, our training dataset may encompass multiple domains with significant disparities, and the data volume in the source domain may far exceed that in the target domain. These factors can potentially impact the effectiveness of domain classifiers, rendering the model less adaptable. In such circumstances, we assign distinct indices to each dataset within a scenario and task the domain classifier with predicting these indices, rather than solely distinguishing between the source and target domains. This approach enhances the feature extractor's ability to capture common (domain-invariant) features across diverse scenarios.
\subsection{Pretraining and Finetuning}
\begin{figure*}[t]
\begin{center}
\subfigure[Chicago]{\includegraphics[width=0.24\linewidth]{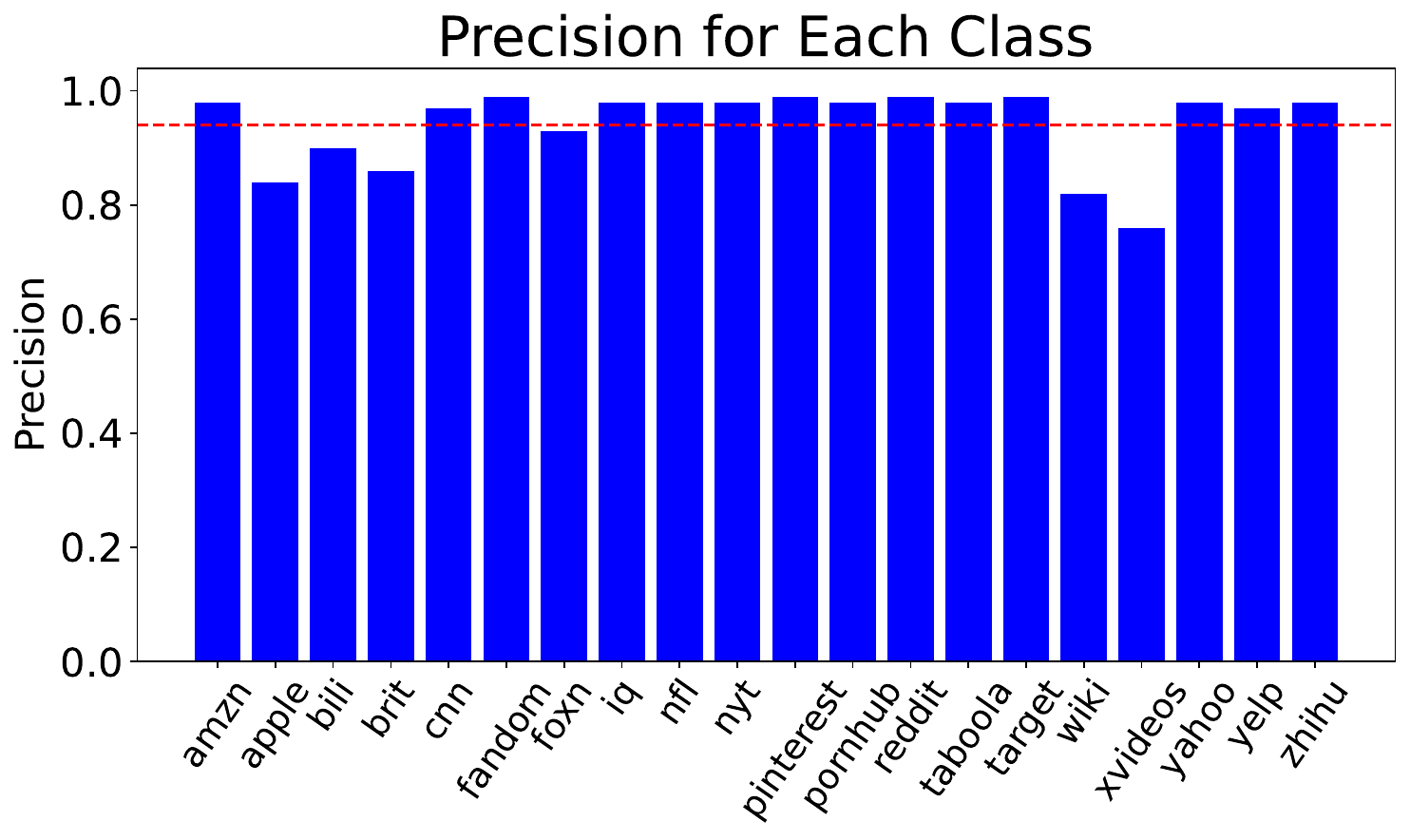}
\includegraphics[width=0.24\linewidth]{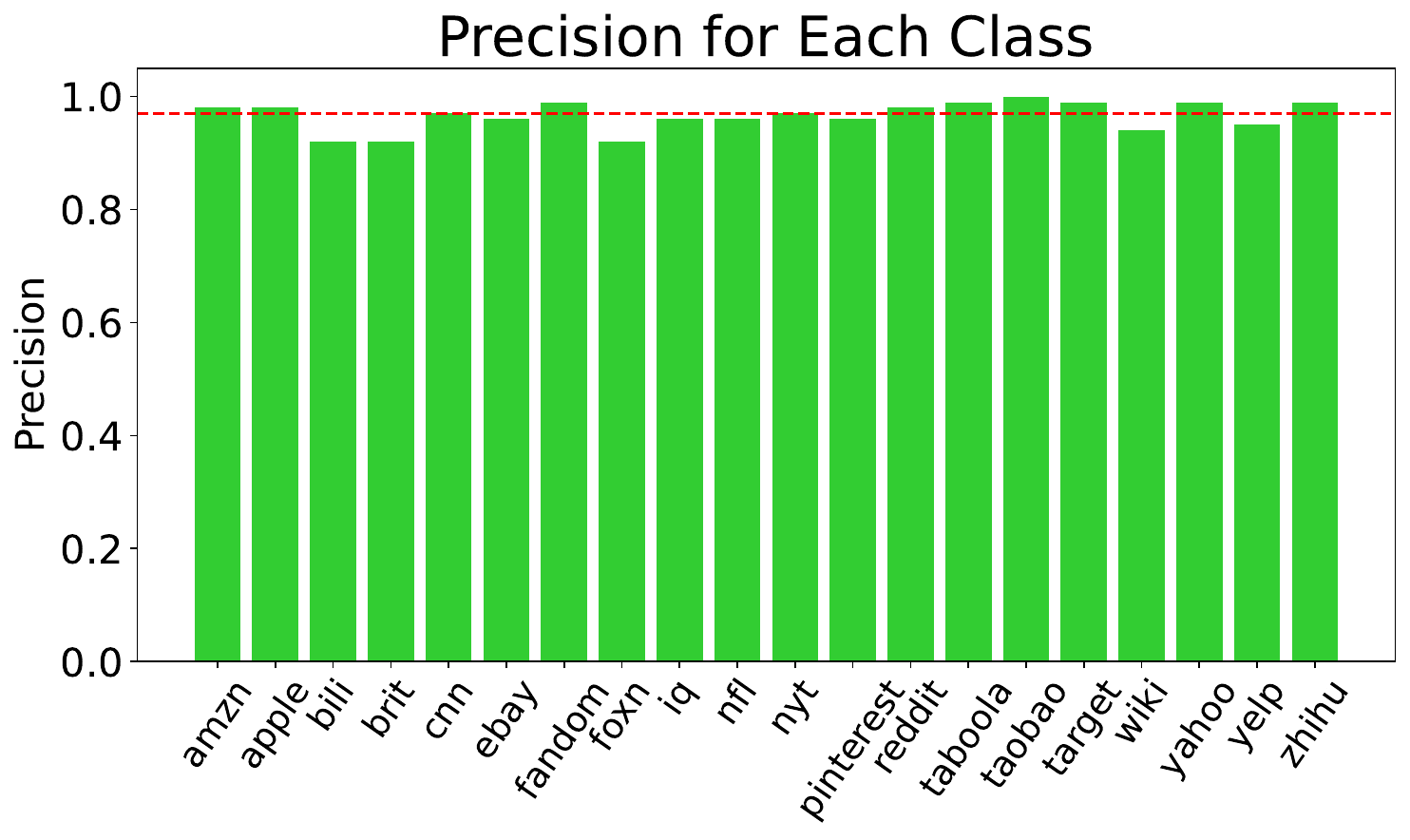}}
\hfill
\subfigure[Miami]{\includegraphics[width=0.24\linewidth]{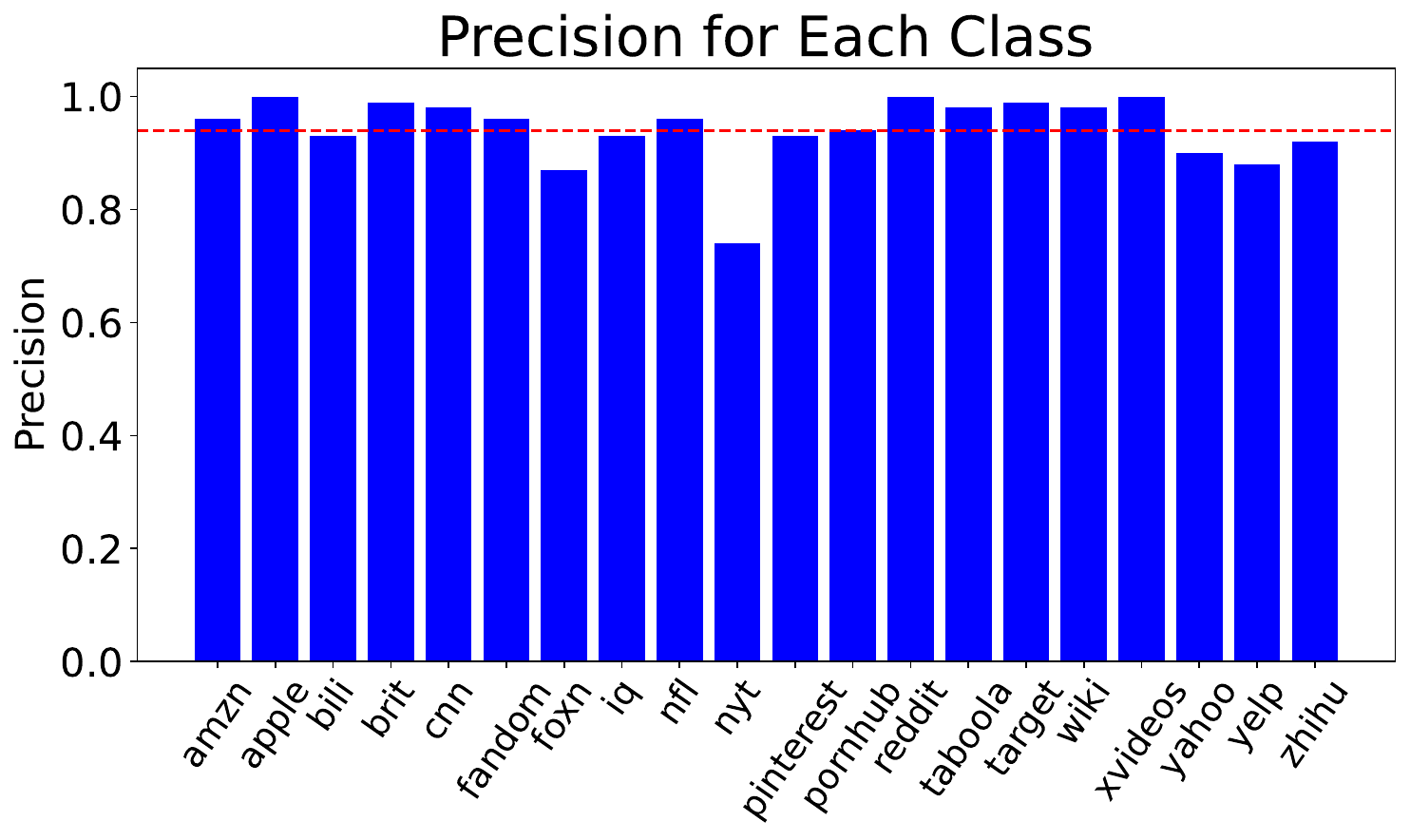}
\includegraphics[width=0.24\linewidth]{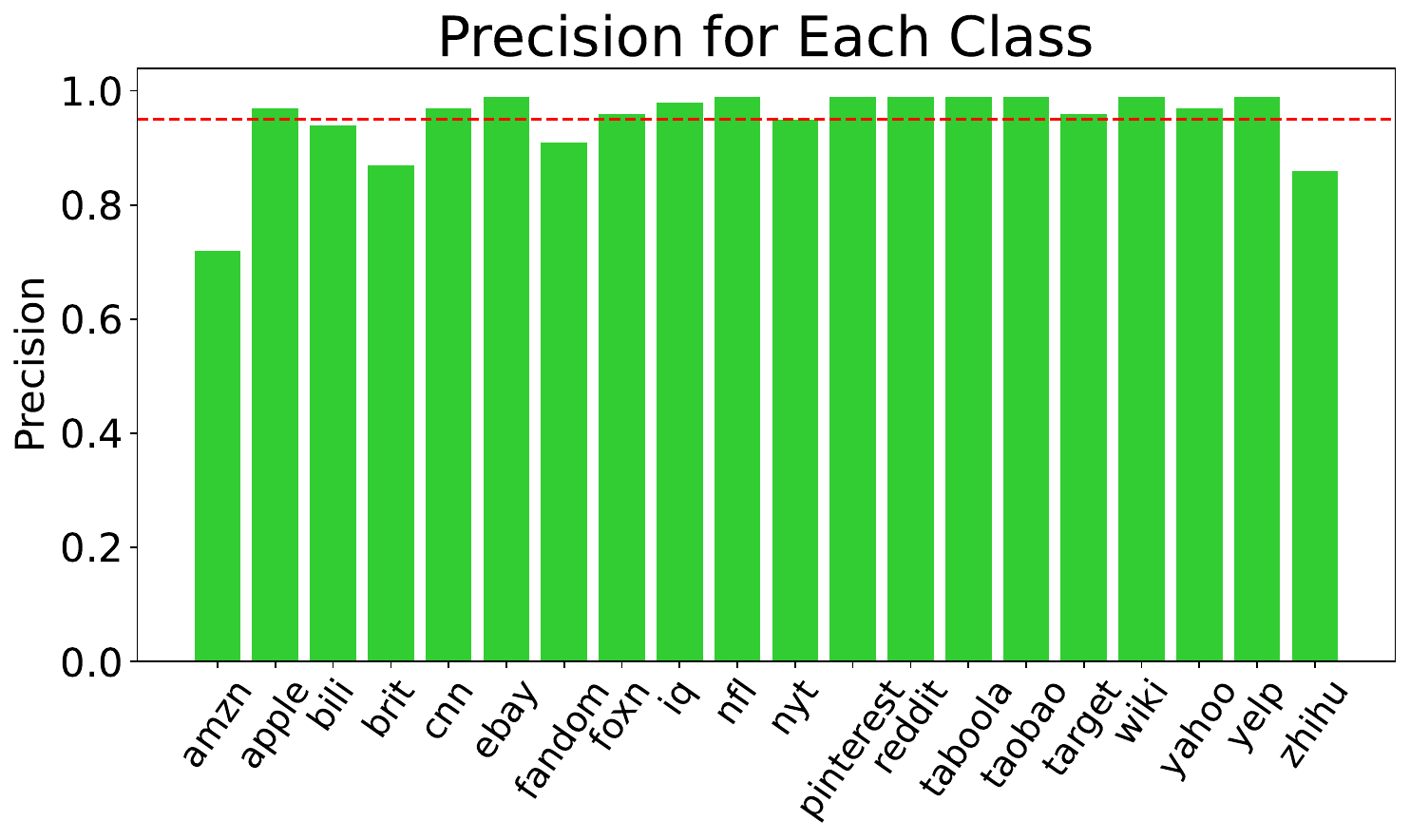}}
\hfill
\subfigure[London]{\includegraphics[width=0.24\linewidth]{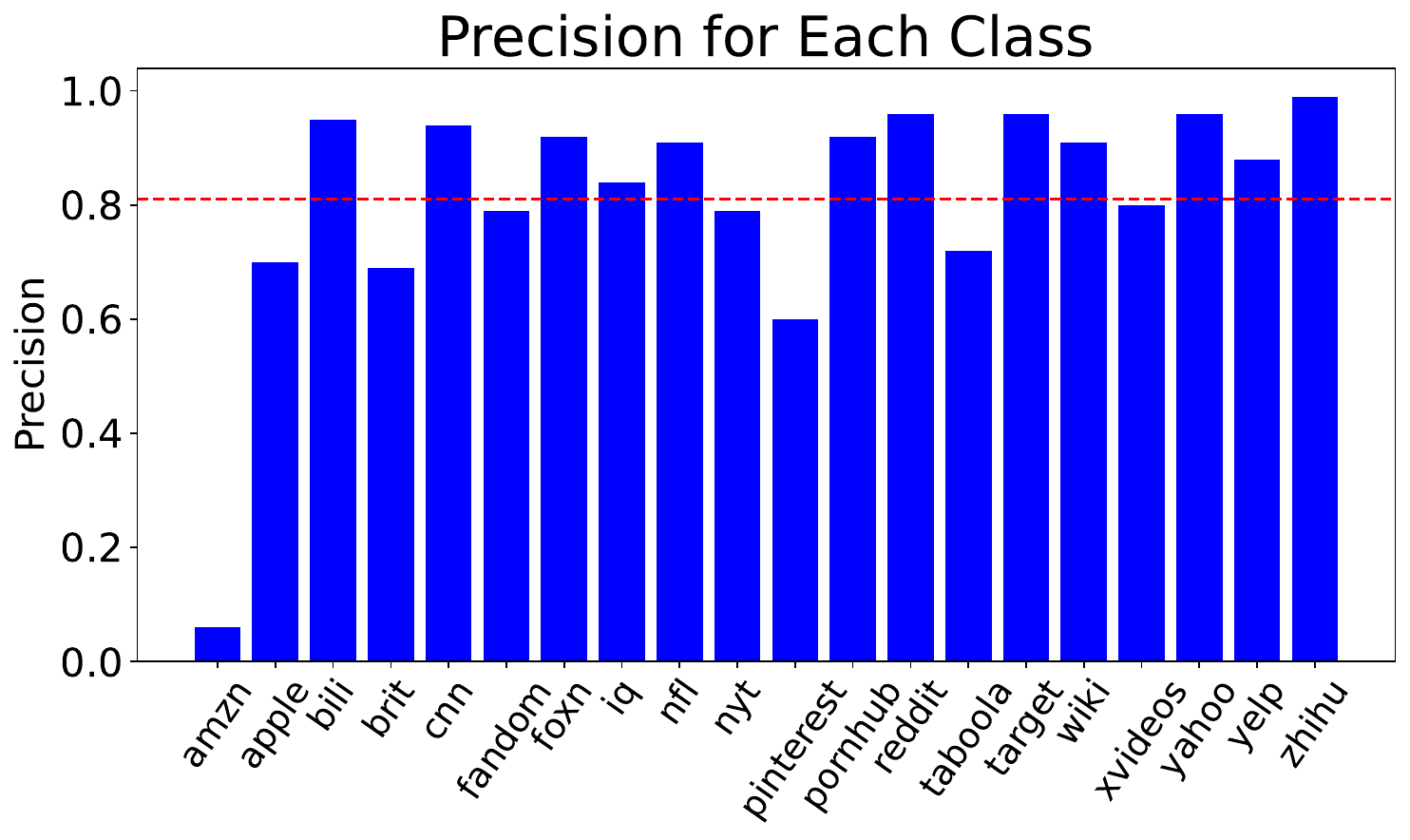}
\includegraphics[width=0.24\linewidth]{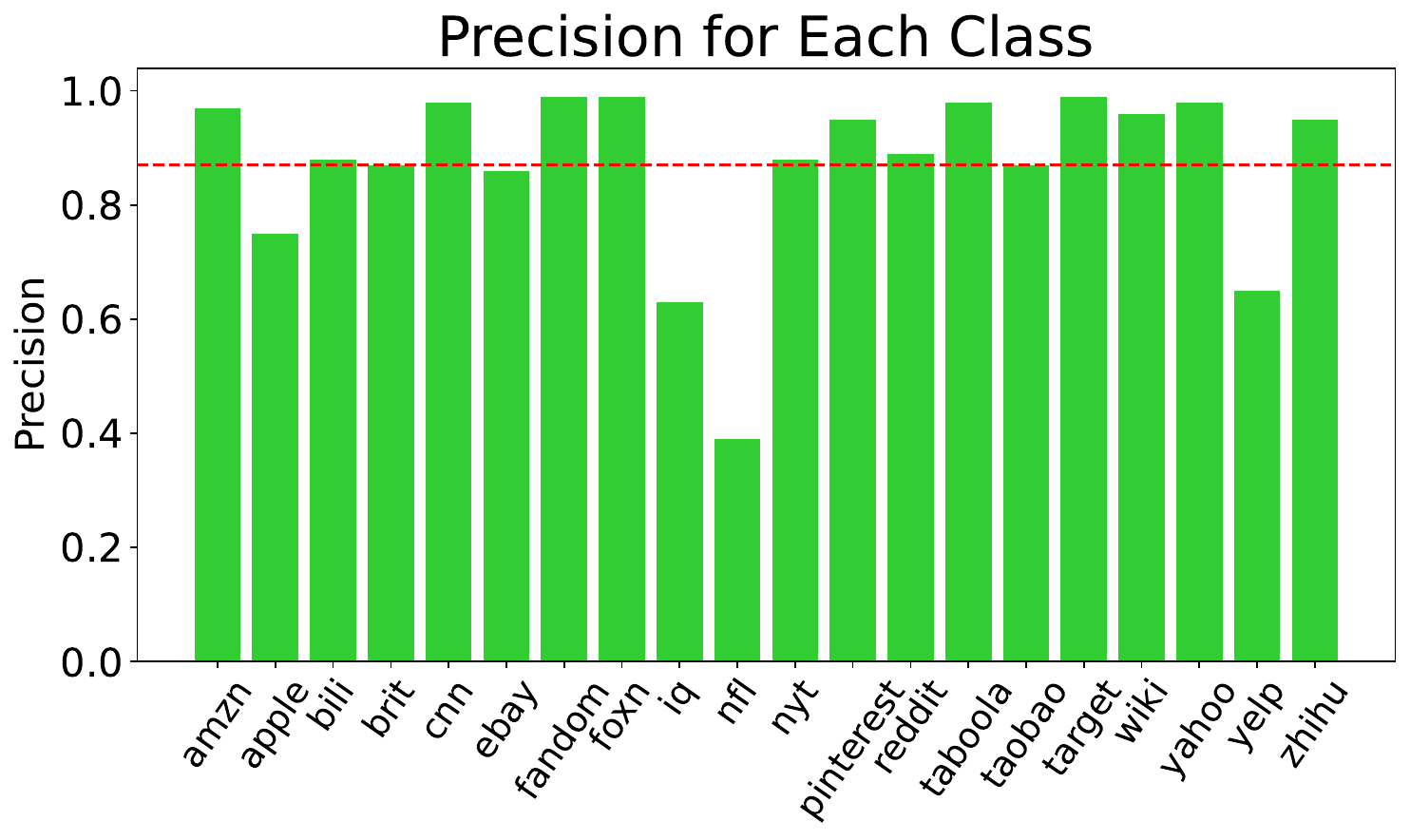}}
\hfill
\subfigure[University]{\includegraphics[width=0.24\linewidth]{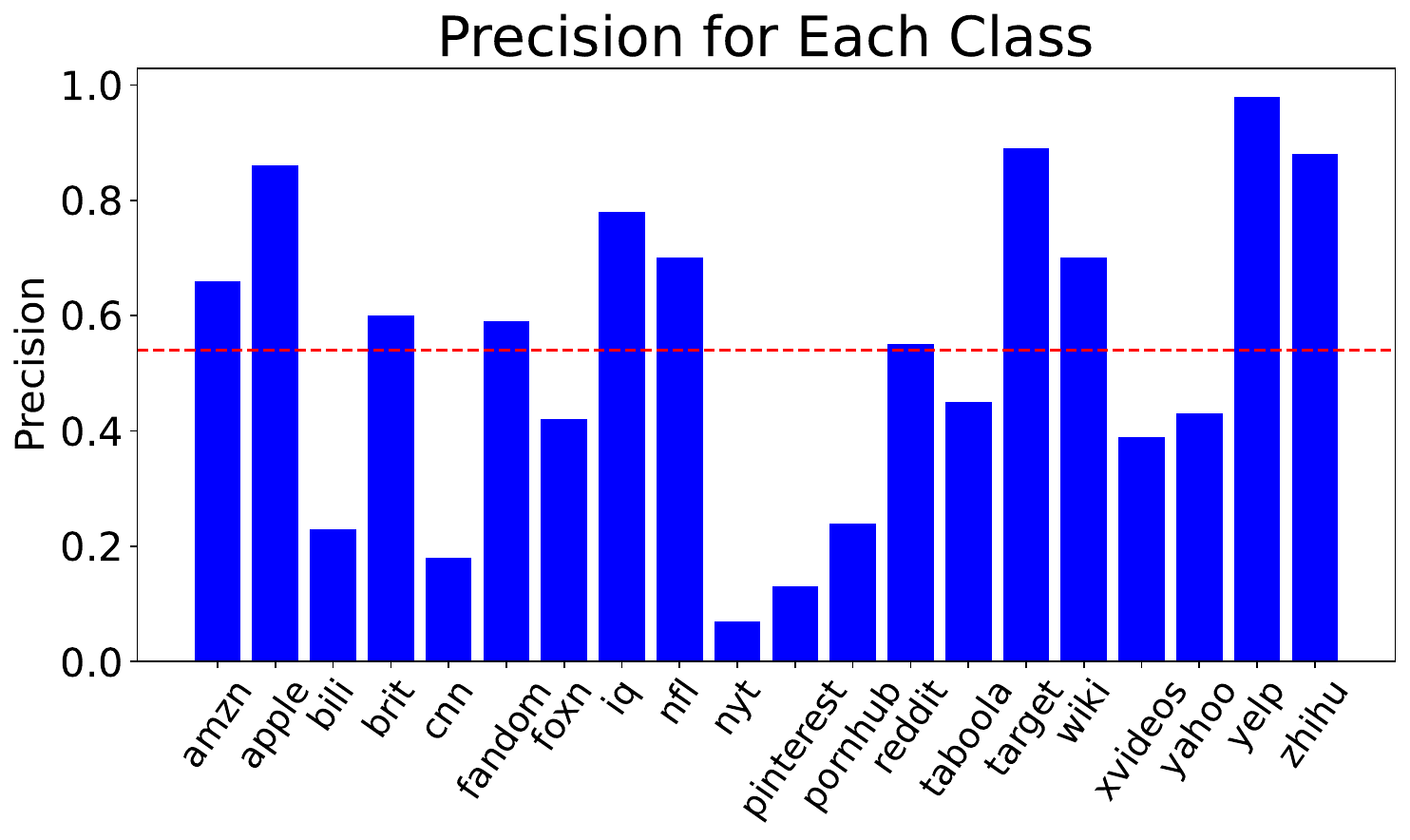}
\includegraphics[width=0.24\linewidth]{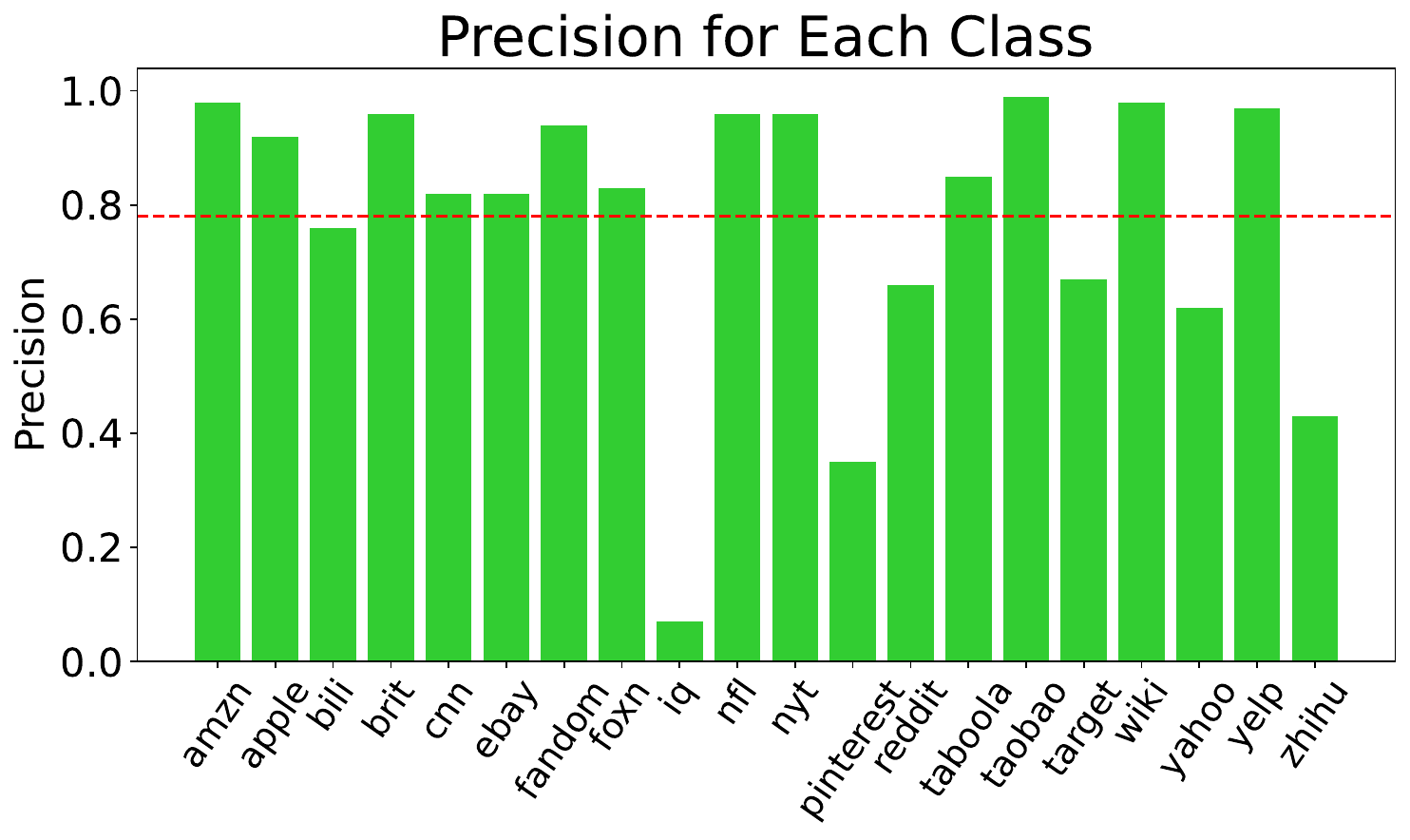}}
\hfill
\caption{\label{fig:domain} Training on multiple locations and testing on an unfamiliar location with (green) and without (blue) domain adaptation.}
\vspace{-0.3cm}
\end{center}
\end{figure*}

For the task of classifying twenty different websites, we combined datasets originating from diverse locations. For each location of the websites, we collected two binary files, each containing over ten million packet flow data. As a consequence, we extracted one file from websites located in all locations except the target one for training purposes, and the other file was designated as the testing dataset. This approach also results in each label corresponding to multiple data flow information from different locations. This approach aimed to train a deeper CNN model, WFNet Large. The intention was for this model to effectively capture distinctive features unique to various websites within larger datasets. Consequently, when dealing with data from unfamiliar locations, we opted to freeze the parameters of the earlier convolutional layers and fine-tune the parameters of the fully connected layers using a relatively small amount of data.

%% file: text/600_evaluation.tex
\section{Evaluation}
\label{sec:experiments}

This section presents our evaluation of WFNet's performance. We construct three attack scenarios: a) the attacker has precise knowledge of the environment used by the victim when browsing websites; b) the attacker knows the range of environments the victim may use for website browsing; c) the attacker has no prior knowledge of the environment in which the victim browses websites. We utilize the WFNet model described in Section \ref{sec:modelstructure} as our foundational attack model and demonstrate its accuracy across these three attack contexts. Subsequently, we investigate the impact of increasing the number of websites and the number of training samples on accuracy.

Section \ref{sec:time} discusses the variations observed when collecting data from the same location but at different times. In Section \ref{sec:da}, we evaluate the enhancements to Scenario C attacks provided by our proposed transfer learning algorithm. We then showcase the results of using single data features—packet interval time or packet size. Although accuracy is reduced, the outcomes remain viable, indicating that each of these features harbors a wealth of site-specific information. Finally, we propose several feasible defense mechanisms against seamless website fingerprint attacks and test their effectiveness. 

\begin{figure*}
\begin{center}
\includegraphics[width=1\linewidth]{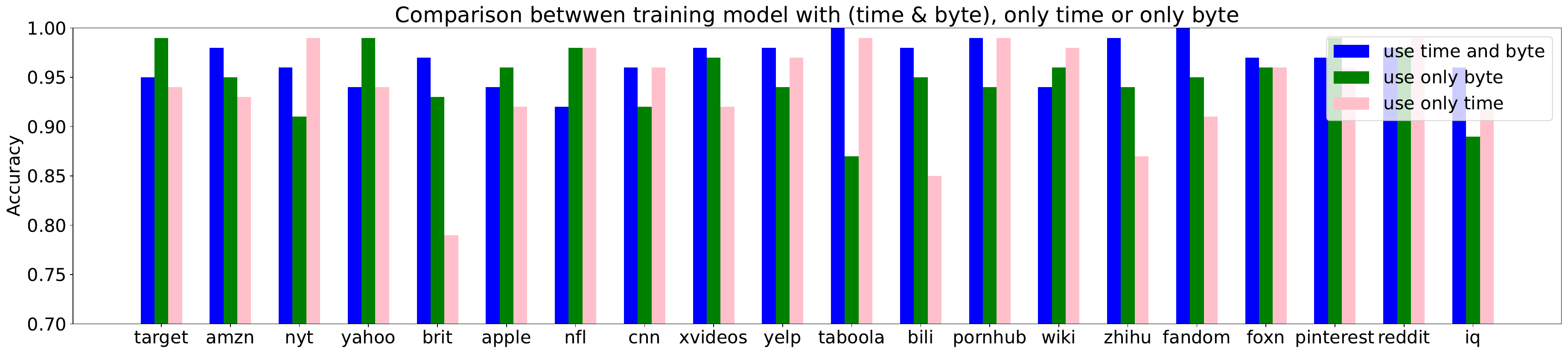}
\caption{\label{fig:only} Accuracy when using jitter, packet size, or both. }
\end{center}
\vspace{-0.6cm}
\end{figure*}
\subsection{Single Location}

Our initial investigation focused on the precision of seamless website fingerprint attacks within a specific environment, premised on the attacker's advanced knowledge of the victim’s device, service provider, and geographical location. The attacker replicates packet traffic for targeted websites using analogous devices in this environment to pre-train the attack model. This experimental setup was replicated across eight distinct datasets, using half of the data (20,000 instances per website) for training, with the remaining divided equally for validation and testing purposes. The validation set's role is to gauge training accuracy, whereas the testing set assesses the WFNet performance post-training.

Precision, defined as the true positive rate, reflects the proportion of correct website identifications among all positive identifications. Meanwhile, accuracy represents the overall rate of correct predictions. Illustrated through a red dashed line, the average accuracy conveys the cumulative performance across all websites, addressing the reliability of the WFNet as a predictor. In subsequent sections, we visually represent the precision of individual websites through bar graphs, with overall accuracy continued to be indicated by a red dashed line.

To verify whether attackers can effectively gather enough information for website-level classification in a seamless fingerprinting environment, we designed two closed-world tasks within a university dataset to reassess the capabilities of three mainstream WFP attack models. According to Table ~\ref{tab:comparison}, the CUMUL model, leveraging SVM classification, and the DLWF model, utilizing a basic deep learning framework, both yielded accuracy rates below 30\% for classifying 10 websites. This rate significantly dropped to approximately 20\% upon increasing the number of websites to 20. In contrast, the Triplet Fingerprint algorithm managed a 75\% accuracy for the 10-website task, though it too faced a notable decrease in accuracy with an increased number of classifications. Given our dataset provides a substantially greater amount of data per class than traditional collections, we inferred that the scope of conventional attack models is inadequately small, lacking the necessary stability and generalization capability for such intricate datasets. Models from other ML domains, like ChatGPT and ResNet, have also demonstrated that deep learning models with a higher number of parameters can significantly improve in stability and generalization when trained with sufficient data. Thus, we opted to increase the depth of our CNN model, employing the WFNet Base and WFNet Large described in section 3.4 for testing. We discovered that an escalation in model size notably enhanced its stability, achieving over 90\% accuracy in both 10 and 20 classification tasks without a discernible decline in accuracy as the number of labels increased.

We next evaluate the applicability of this attack strategy across various network environments. Figure \ref{fig:single-location} demonstrates that, given a substantial training dataset, our model consistently achieves an accuracy rate above 90\% across all datasets. This finding emphasizes the potential for accurate attacks when the attacker possesses detailed knowledge of the victim's network settings, underscoring the considerable risk associated with WF attacks under conditions where attackers can emulate the victim's browsing environment closely. Notably, each input vector comprises 500 packets, indicating that in our rigorously defined seamless website fingerprinting scenario, brief packet samples suffice to disclose a wealth of personal browsing data, highlighting the inherent privacy vulnerabilities within network traffic patterns.

\subsection{All Traffic Sets}
\begin{figure}[t] 
    \centering
    \includegraphics[width=8cm]{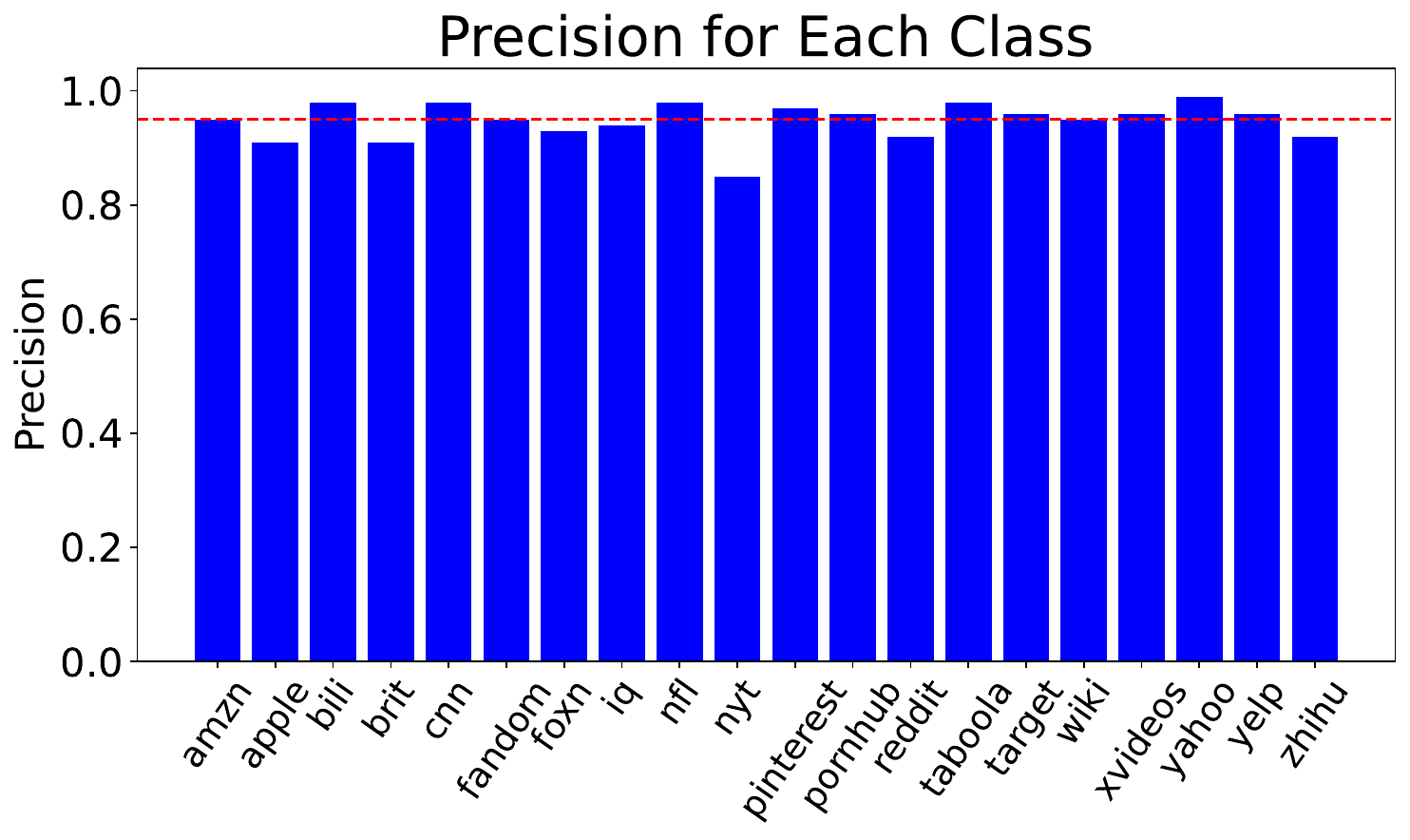}
    \caption{\label{fig:Overview} Precision of each website using all traffic sets for both training and testing. The red dashed line shows 95\% overall accuracy.}
\end{figure}

Subsequently, we explored the potential for attackers to conduct predictions across multiple environments simultaneously. We operated under the assumption that attackers are aware of the range of environments a victim might use for web browsing, such as the types of devices or their locations, with the aim of developing a model capable of executing attacks across various environments.

In our experiments, we amalgamated the traffic from 20 websites and all 8 locations mentioned in Table~\ref{table111} into a single dataset. To thoroughly assess the model's performance across each environment, we evenly distributed the eight datasets between the training and test sets. Figure~\ref{fig:Overview} displays the precision for each website and the collective accuracy when all traffic datasets are combined. In this evaluation, our model achieved an overall accuracy of 95\%. Notably, only one website, The New York Times, recorded a precision below 90\%. This indicates that with sufficient data for preliminary training, attackers could feasibly develop a model with robust generalization capabilities, enabling seamless attacks on victims across multiple scenarios.

\subsection{Different Locations}
\label{sec:diff}

While amalgamating data from various sources improves accuracy, it doesn't elucidate the distinct contributions of different environments, nor does it address the implications of collecting training data in one environment and applying it in another. As previously discussed, the real-world browsing environment is affected by continuous factors, making it exceedingly difficult for attackers to identify all possible network environments a victim might use, due to the prohibitive cost and complexity. This section delves into the generalization capability of attack models in novel environments, operating under the assumption that attackers cannot pinpoint the victims' specific browsing contexts. This necessitates attackers to initially train their models in familiar environments before testing their applicability in new, uncharted territories.

To assess the influence of network and CPU environments on model performance, we analyzed the precision achieved when training and validation data originated from one location, while the testing data was sourced from another. Table \ref{tab:locations} presents how the training set location significantly impacts accuracy, offering a matrix that maps out classification accuracies across various training and testing environment pairings. Accuracy is notably high along the matrix diagonal, where training and testing environments match, showcasing the effectiveness of environment-specific training. Off-diagonal entries reveal substantial accuracy declines, highlighting challenges in cross-environment generalization. For example, training with data from \vultr-London and classifying traffic from \optimum home 2 resulted in only 23\% of traffic samples being correctly classified, illustrating the stark accuracy falloffs when transitioning across cloud providers.

Furthermore, the "WFNet Base source only" data presented in Table \ref{tab:target-source} reveals the model's performance in unfamiliar settings post-training across seven distinct target environments. Diversifying training data does indeed bolster the model's ability to generalize beyond a single training context. Yet, the gains in performance are somewhat muted. Directly targeting users in a \university setting, for example, sees the model's accuracy halving to just 50\%. This demonstrates the significant real-world challenges website fingerprint attacks face, underlining the substantial practical hurdles in applying these attacks across diverse network environments.

\subsection{Number of Websites}
\label{sec:numbers}
\begin{table}[t]
\newcommand{\NA}{---}
\caption{\label{tab:volume} Accuracy trend as the number of training samples increases.}
\scalebox{0.68}{
\begin{tabular}{@{}l|llllllll@{}}
\toprule
Number of sample vectors used for training                             & 1000 & 2000 & 5000 & 10000 & 20000 & 40000 & 60000 & 100000 \\ \midrule
Mix of all locations                                 & 0.27 & 0.7  & 0.83 & 0.85  & 0.91  & 0.93  & 0.94  & 0.95   \\
WFNet Base at \university                            & 0.29 & 0.56 & 0.8  & 0.87  & 0.91  &   \NA    &   \NA    &   \NA     \\
Pretrain on other location and fine-tune at \university & 0.78 & 0.83 & 0.87 & 0.89  & 0.92  &   \NA    &   \NA    &  \NA      \\ \bottomrule
\end{tabular}}
\end{table}

A key effect to explore is how accuracy scales with the number of websites. One problem with CNNs is they
sometimes collapse if the data sets introduce too similar patterns as the number of classes increases. For example, if websites were
``copy cats'' of one another in some matter, e.g. all using the same template, it might result in indistinguishable traffic patterns. Table~\ref{tab:number} shows the results from continuously collected traffic from the \university location. Within this environment, we have amassed
traffic from 40 popular websites, which include several websites of similar categories, such as travel websites (e.g., Airbnb, Expedia, Booking) and news websites (e.g., CNN, BBC, The New York Times), among others. As illustrated in the first row of Table~\ref{tab:number}.
These results show a very stable accuracy as the number of websites increases from 10 to 40, with the average
accuracy stabilizing at about 91\%. We can thus have relatively high confidence that for the distinctive websites in
our set, adding additional websites will not reduce accuracy. 

Another pivotal factor influencing the robustness of the model pertains to the quantity of samples utilized for training. Table~\ref{tab:volume} demonstrates the correlation between the number of sample vectors utilized and the resultant accuracy. It's important to note that each sample vector, whether for training or classification, comprises 1000 elements, split evenly between jitter times and packet sizes, input as pairs. The data indicates that once the count of vectors per site surpasses approximately 50,000, accuracy plateaus at a mid-90\% range. Consequently, to achieve training sets that yield mid-90\% accuracy, approximately 25-30 million packets per website are requisite.


\subsection{Aged Traffic}
\label{sec:time}

Websites are in a constant state of flux, regularly updating their layouts and introducing new features. This dynamism prompts an examination of a trained network's resilience to temporal changes spanning several months. A pivotal question we address is the extent to which accuracy deteriorates over time. The second row of Table~\ref{tab:number} elucidates the effect of utilizing a training set that is nine months old. For this experiment, the model was trained on a dataset from the University, gathered nine months prior (December 2022), and subsequently tested on a later University dataset (August 2023) across classification tasks for 10, 15, and 20 websites. The findings reveal that relative to concurrent training and testing on the newer dataset, the foundational WFNet model witnesses an accuracy decline of roughly 10 to 15 percentage points. Intriguingly, as the task complexity increases with more websites, the trend in accuracy does not consistently worsen. Moreover, the application of domain adaptation techniques moderates this drop in accuracy, maintaining it within a five to eight percent margin.

\begin{table}[t]
\centering
\caption{Accuracy as the number of websites increases, for both aged training traffic and domain-adapted WFNet. Current-only shows training and testing on the current University data set. Old-only shows training on the old University data set and testing on the current University data set. }
\label{tab:number}
\newcommand{\NA}{---}
\begin{tabular}{llllllll}
\toprule
Number of websites                                                                      & 10   & 15   & 20     & 25   & 30   & 35   & 40  \\ \midrule
WFNet (Current-only)                                               & 93   & 92   & 91.3   & 90   & 90   & 91   & 91  \\
WFNet (Old-only)                                                   & 80   & 78   & 79     &   \NA   &   \NA   &  \NA    &  \NA   \\
Domain adapted WFPNet (Old-only)                                          & 87   & 88   & 85     &   \NA   &   \NA   &  \NA    &  \NA   \\ \bottomrule
\end{tabular}
\end{table}

\subsection{Domain Adaptation}
\label{sec:da}
ominantly encounter Scenario C, where they lack foreknowledge of the victim's browsing environment and timing. This constraint prevents them from simulating and acquiring adequately labeled data in advance. Variabilities in external environments and timing contribute to dataset shifts, compromising the attackers' efficiency in unfamiliar settings. This section evaluates the efficacy of two strategies attackers might employ, alongside corresponding optimization algorithms for transfer learning.

In the first scenario, once attackers identify a victim, they can swiftly generate a minimal amount of labeled ground truth data for the websites in question using devices akin to those of the victims within similar environments. This process enables attackers to fine-tune their existing attack models with this sparse data to enhance their attack's effectiveness. Our experiments directed the attack models to train on datasets excluding the university dataset (with 100,000 instances per class) and later fine-tuned with varying sample sizes from the university training set. As demonstrated in the last row of Table 6, fine-tuning with 1,000 instances per class escalated the model's effectiveness from 50\% (with no fine-tuning) to 78\%. In contrast, constructing a new model from scratch with the same amount of instances achieved merely a 23\% accuracy rate. This signifies that although attack models may not be directly deployable in new environments, they can still reach substantial levels of attack precision in significantly reduced timeframes (one-tenth or even one-twentieth of the original time).

In the second scenario, aimed at rapidly acquiring victims' browsing information for swift deployment and attack (a more harmful attack scenario), attackers can utilize domain adaptation algorithms. These algorithms seek domain invariants between the model's original training set (labeled) and the new dataset (unlabeled), employing these invariants to negate the effects of varying environments or timings. Figure \ref{fig:domain} illustrates that domain adaptation generally boosts average precision and accuracy, albeit with a decrease observed in a few websites. For instance, while the overall average precision and accuracy for the London location saw an increase, precision for several sites dropped. The figure also reveals that the impact of domain adaptation is relatively minor when overall accuracy is high, as seen in the Chicago site. Conversely, at the University site, where overall accuracy typically is low, domain adaptation displayed the most pronounced improvement, elevating the average accuracy from below 60\% to nearly 80\%. Hence, we deduce that domain adaptation proves to be an effective strategy, particularly when aiming to enhance average accuracy as the performance benchmark.

\begin{table}[]
\centering
\caption{\label{tab:target-source}Source-only means a model trained using multiple locations and tested on a new, unfamiliar location. Target-only represents model trained and tested only on a single location.}
\scalebox{0.6}{
\begin{tabular}{@{}l|cccccccc@{}}
\toprule
\hspace{1.0cm} Model (Task)                           & Chicago & London & Miami & University & Toronto & San Francisco & Homecable 1 & Homecable 2 \\
WFNet Base (Source-only)         & 92      & 70     & 87    & 51         & 86      & 88            & 0.68        & 0.77        \\
Domain adaptation (Source-only) & 97      & 87     & 95    & 78         & 88      & 91            & 0.74        & 0.81        \\
WFNet Base (Target-only)           & 97      & 92     & 95    & 90         & 97      & 96            & 90          & 89          \\ \bottomrule
\end{tabular}}
\end{table}


\subsection{Only Jitter and Size}
One question is what is the accuracy if we used only jitter or only packet sizes? The first reason is that
a classification system using one would use less resources than one using both,
If reasonable accuracy can be obtained with just one of these two features it would make training data collection
easier, and more accessible to middleboxes and routers. 
Second, if Web sites' patterns
are visible in a single feature, it means there is a very powerful causality between web design and traffic
patterns that may be difficult to obscure. 

Figure~\ref{fig:only} compares the accuracy when using both features, as well as only jitter and size for 50K sample vectors,
When exploring a single feature, we set the other feature values to all zeros in the 1000-element vector. 
The figure shows a very slight decrease in average accuracy, from 95.6\% for both to 92\% for jitter alone and 93\% for size alone.
This result shows that both packet jitter and size reflect much of the inherent structure of the website.

%% file: text/700_protection.tex
\section{Protecting Privacy}
\label{sec:protection} 
\begin{figure*}[t]
\begin{center}
\subfigure[Inflation]{\includegraphics[width=0.49\linewidth,height=5cm]{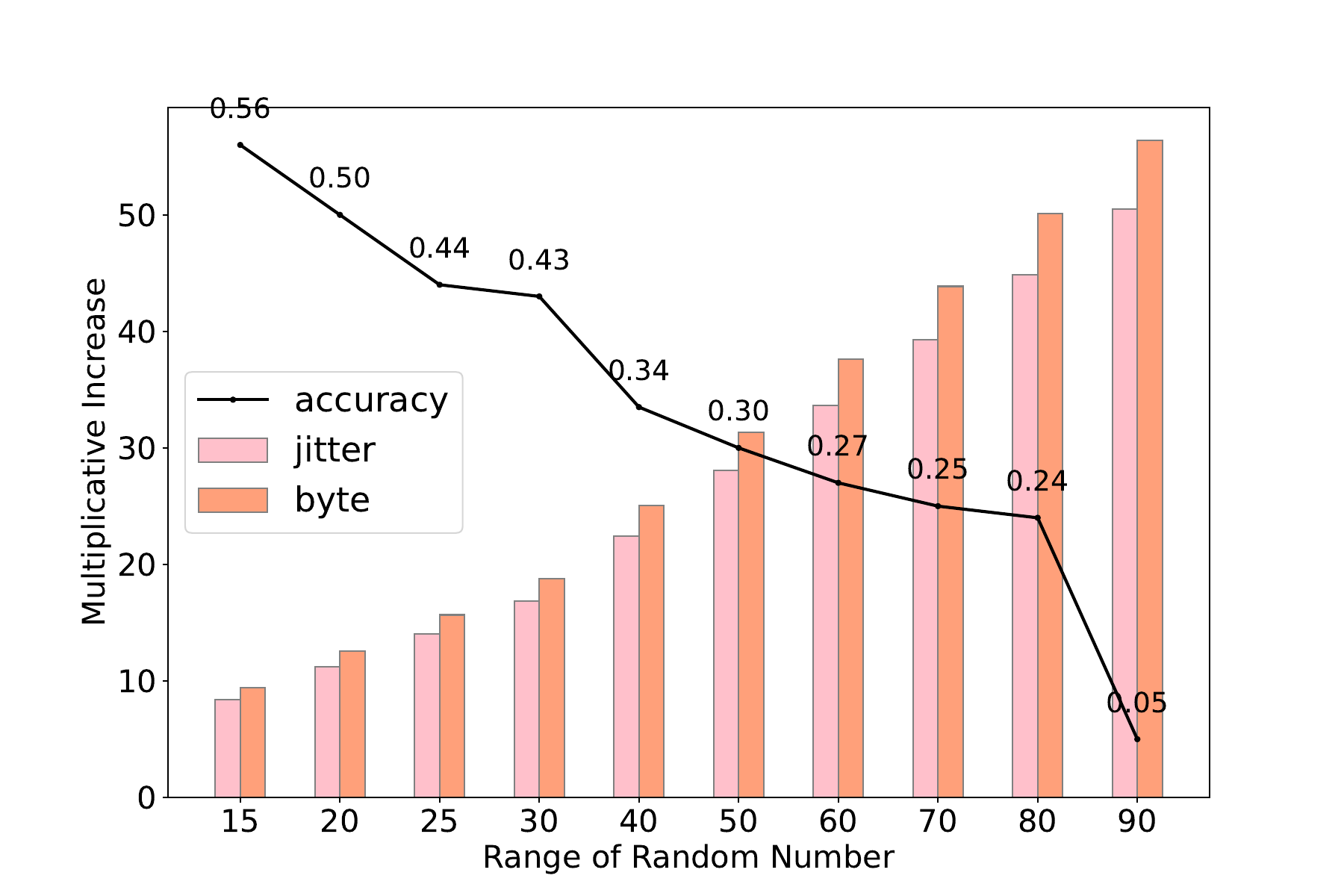}}
\hfill
\subfigure[Active Injection]{\includegraphics[width=0.49\linewidth,height=5cm]{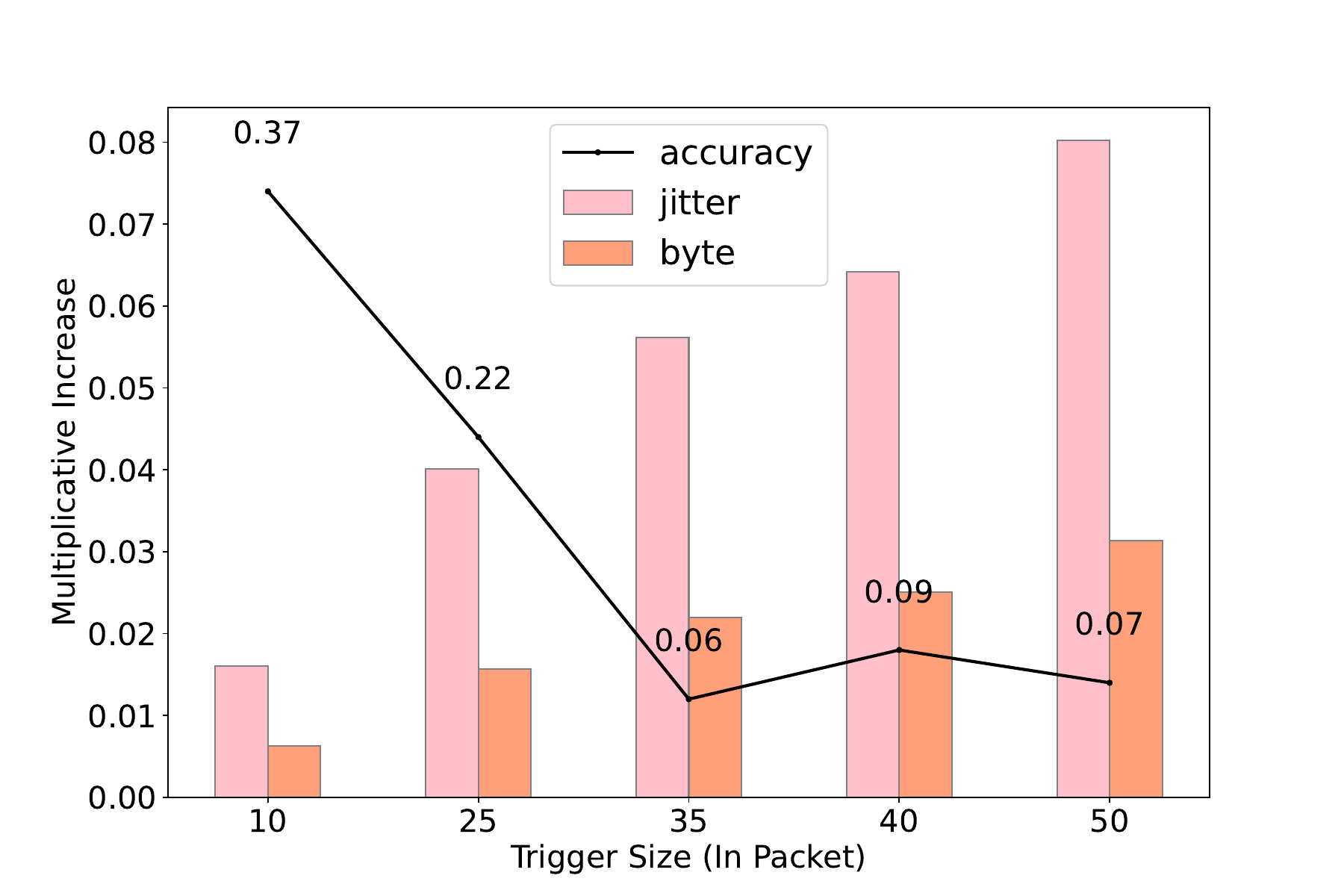}}
\caption{\label{fig:privacy} Effectiveness of Evasion by adding randomization. Figure (a) shows the impact of Inflation, while
Figure (b) shows the impact of Active Injection.}
\end{center}
\vspace{-0.6cm}
\end{figure*}

This section shows how to improve privacy protection by using two evasion techniques that add randomization
to contaminate traffic features. The first, {\em Inflation}, only alters existing packets. The second, {\em Active Injection} actively adds packets. We first
describe these techniques and then evaluate their performance by measuring the model's resulting drop in accuracy. 

\subsubsection{Inflation}
Inflation adds a randomized extra delay to the jitter of a packet or pads the packet with a random amount of bytes. 

The value of such an approach is that it minimizes protocol changes and is more straightforward to run in middleboxes such as routing nodes and switches. We introduce a random number between 0 and $X$ for each data packet's time interval and size. Here, $X$ is computed based on the average values of all interval times or sizes in the observed traffic, multiplied by a coefficient, denoted as $a$ to determine the specific value of $X$. In our experiments, we consider ten different $a$, namely [15, 20, 25, 30, 40, 50, 60, 70, 80, 90], and apply this random algorithm to both the training and testing datasets. The goal of this method is to disrupt data interception or analysis attempts based on predictable transmission patterns, thereby enhancing data transmission security and bolstering resistance against privacy infringements.

\subsubsection{Active Injection}

 Motivated by the vulnerabilities associated with backdoor attacks, we present an approach involving the insertion of trigger points that inject a randomized set of extra packets, aimed at deliberately contaminating both training and testing datasets. This intentional contamination significantly diminishes the accuracy of Machine Learning (ML) models in their predictions of website outcomes. Both the transmitting and receiving parties incorporate trigger points into the data stream after transmitting a pre-defined number of packets. This sequence of packets is meticulously crafted to exhibit a distinct and predetermined distribution, such as a series of 15 packets each with a length of 30 bytes. 
 Due to the protocol's random selection of triggers from the trigger pool and their insertion into the packet stream, along with the temporal variation in trigger usage on a daily or weekly basis, it leads to a situation where the model might encounter different trigger points inserted into the data associated with the same label during both training and testing phases. This incorporation of triggers also leads to the emergence of highly noticeable, specific, and controllable features within the packet transmission flow. Our experiments demonstrate that as the trigger size increases, ML models increasingly rely on triggers for website predictions, establishing a correlation between website labels and inserted triggers. Furthermore, intentionally altering the inserted triggers can substantially alter the ML model's label predictions, thereby safeguarding user privacy.

In our experiments, we explore varying triggers, namely $2\%$, $4\%$, $7\%$, $8\%$ and $10\%$ of the original number of packets (i.e., inserting 10, 25, 35, 40, and 50 packets every 500 packets). Within these inserted triggers, we maintain a constant size between 50-250 bytes per packet within a trigger, as different triggers use different-sized packets. The inter-arrival time of each packet was a random number between 0 and the average jitter. Figure 2 demonstrates that this approach, with an increase of less than $8\%$ in additional time and data volume, consistently reduces the model accuracy to single digits. This outcome significantly outperforms the direct addition of latency and size to existing packets. \ref{sec:experiments} show more detail about its performance.

\subsection{Evaluation}

Figure~\ref{fig:privacy}(a) shows the impact of Inflation, and Figure~\ref{fig:privacy}(b) shows the impact of Active Injection.
Figure (a) shows the decrease in accuracy when using inflation to add jitter and extra bytes to existing traffic.
The X-axis shows the maximum number of standard deviations of jitter or bytes to add to the packet. The resulting increased
average absolute increase in jitter or bytes from the added randomization is shown on the Y-axis. The dark black line shows the resulting accuracy.
A router could perform jitter inflation by adding a random delay to a packet. Padding TCP packets is more complex, but is
relatively straightforward with split-TCP strategies that interpose on the TCP stream.

Figure~\ref{fig:privacy}(a) shows that reducing accuracy using randomized inflation imposes costs that are likely to
be untenable for a web browsing application. Getting the classification accuracy to 5\% requires setting the randomization
maximum for uniform distribution to be 90 standard deviations over the average standard deviations for jitter and
size. This adds an average of $50x$ delay to the packets and increases the total data size to over $50x$ of the original
volume of data sent. 

Figure~\ref{fig:privacy} (b) shows the effectiveness of active injection. The X-axis shows the trigger size, that is the number of extra, fixed-sized
packets injected (the trigger), The Y-axis shows the amount of extra total delay and multiple bytes over the base case of no trigger.
The black line shows the resulting accuracy. The figure shows that when adding a trigger size of 35 packets, website identification accuracy using the trained
WFNet drops to 6\%. The excess average delay only increases by 6\% and the extra size of the traffic only increases by
2\%.

%% file: text/900_Conclusion_Future_Work.tex
\section{Discussion and Conclusion}
\label{sec:conclusion}

In this work, we showed how the ability of deep neural networks to find patterns in large data sets impacts the field
of computer networks, in particular the privacy implications of web browser traffic at the TCP level. 
Our research demonstrates that it is possible to identify individual websites by only using the packet inter-arrival time and length,
thus providing a privacy risk even when using masking techniques such as encryption and VPNs. 
We developed a novel CNN, WFNet, that obtained high accuracy. We showed this high accuracy required collecting traffic traces from a diverse set of environments, including
several cloud computing providers, a well-connected university server, and cable modems from private homes. We showed how domain adaptation
reduces the impact of the network environment on the CNN accuracy, allowing for good accuracy from locations for which there is no
training data. 

While our results only examined TCP streams running over HTTPS, we suspect that newer protocols such as QUIC and HTTP3 over IPv6
will have similar results. If it is true that website structure results in identifiable patterns that can be found by CNNs at the HTTPs/TCP levels,
then such patterns will also be visible in QUIC and derived protocols.

We found that although using both jitter and packet size resulted in high classification accuracy, using either alone only 
resulted in a few percentage accuracy drops. Our work provides strong evidence that website structures, such as which
templates are used, or how advertisements are placed, likely result in identifiable patterns at the TCP and IP level.
An important question our work raises is understanding the connection between website structure and the
patterns are seen at the packet level.

One way to obtain additional privacy may be to copy such website structures, i.e. build a sort of copy-cat site, where the structure
of the website is copied but the content remains different from the original. Future work in this area would start with cloned sites
and alter features to better understand which website structures map to which traffic patterns. 

Finally, our work points to a more long-term strategy of how to preserve privacy from attackers that can observe system-level
timing and size events. Our initial work at obfuscation points to a strategy of allowing protocols to be able to add randomized data
within the protocol in a manner that is not clear to the attacker it is randomized traffic. For example, allowing protocols
to add random packets with null effects that exist only to ensure privacy against machine learning models.
Such protocol enhancements would allow a wealth of privacy-preserving techniques
based on randomized injection techniques to exist without altering the protocols themselves.

%% file: main.bbl
\begin{thebibliography}{10}
\providecommand{\url}[1]{#1}
\csname url@samestyle\endcsname
\providecommand{\newblock}{\relax}
\providecommand{\bibinfo}[2]{#2}
\providecommand{\BIBentrySTDinterwordspacing}{\spaceskip=0pt\relax}
\providecommand{\BIBentryALTinterwordstretchfactor}{4}
\providecommand{\BIBentryALTinterwordspacing}{\spaceskip=\fontdimen2\font plus
\BIBentryALTinterwordstretchfactor\fontdimen3\font minus
  \fontdimen4\font\relax}
\providecommand{\BIBforeignlanguage}[2]{{%
\expandafter\ifx\csname l@#1\endcsname\relax
\typeout{** WARNING: IEEEtranS.bst: No hyphenation pattern has been}%
\typeout{** loaded for the language `#1'. Using the pattern for}%
\typeout{** the default language instead.}%
\else
\language=\csname l@#1\endcsname
\fi
#2}}
\providecommand{\BIBdecl}{\relax}
\BIBdecl

\bibitem{10.1145/3447382}
\BIBentryALTinterwordspacing
I.~Akbari, M.~A. Salahuddin, L.~Ven, N.~Limam, R.~Boutaba, B.~Mathieu,
  S.~Moteau, and S.~Tuffin, ``A look behind the curtain: Traffic classification
  in an increasingly encrypted web,'' \emph{Proc. ACM Meas. Anal. Comput.
  Syst.}, vol.~5, no.~1, feb 2021. [Online]. Available:
  \url{https://doi.org/10.1145/3447382}
\BIBentrySTDinterwordspacing

\bibitem{aminuddin2023rise}
M.~A. I.~M. Aminuddin, Z.~F. Zaaba, A.~Samsudin, F.~Zaki, and N.~B. Anuar,
  ``The rise of website fingerprinting on tor: Analysis on techniques and
  assumptions,'' \emph{Journal of Network and Computer Applications}, vol. 212,
  p. 103582, 2023.

\bibitem{Kaur_2014}
\BIBentryALTinterwordspacing
P.~K. and, ``Web content classification: A survey,'' \emph{International
  Journal of Computer Trends and Technology}, vol.~10, no.~2, pp. 97--101, apr
  2014. [Online]. Available:
  \url{https://doi.org/10.14445%2F22312803%2Fijctt-v10p117}
\BIBentrySTDinterwordspacing

\bibitem{auld_2007_bayesian}
T.~Auld, A.~W. Moore, and S.~F. Gull, ``Bayesian neural networks for internet
  traffic classification,'' \emph{IEEE Transactions on neural networks},
  vol.~18, no.~1, pp. 223--239, 2007.

\bibitem{bissias2006privacy}
G.~D. Bissias, M.~Liberatore, D.~Jensen, and B.~N. Levine, ``Privacy
  vulnerabilities in encrypted http streams,'' in \emph{Privacy Enhancing
  Technologies: 5th International Workshop, PET 2005, Cavtat, Croatia, May
  30-June 1, 2005, Revised Selected Papers 5}.\hskip 1em plus 0.5em minus
  0.4em\relax Springer, 2006, pp. 1--11.

\bibitem{10001054}
T.~Burns, C.~Song, I.~Seskar, J.~Ortiz, and R.~P. Martin, ``A simplified
  machine learning approach to classifying individual websites,'' in
  \emph{GLOBECOM 2022 - 2022 IEEE Global Communications Conference}, 2022, pp.
  6109--6114.

\bibitem{NEURIPS2021_45017f65}
\BIBentryALTinterwordspacing
H.-Y. Chen and W.-L. Chao, ``Gradual domain adaptation without indexed
  intermediate domains,'' in \emph{Advances in Neural Information Processing
  Systems}, M.~Ranzato, A.~Beygelzimer, Y.~Dauphin, P.~Liang, and J.~W.
  Vaughan, Eds., vol.~34.\hskip 1em plus 0.5em minus 0.4em\relax Curran
  Associates, Inc., 2021, pp. 8201--8214. [Online]. Available:
  \url{https://proceedings.neurips.cc/paper_files/paper/2021/file/45017f6511f91be700fda3d118034994-Paper.pdf}
\BIBentrySTDinterwordspacing

\bibitem{chen2019measuring}
T.~Chen, W.~Cui, and E.~Chan-Tin, ``Measuring {ToR} relay popularity,'' in
  \emph{Security and Privacy in Communication Networks: 15th EAI International
  Conference, SecureComm 2019, Orlando, FL, USA, October 23-25, 2019,
  Proceedings, Part I 15}.\hskip 1em plus 0.5em minus 0.4em\relax Springer,
  2019, pp. 386--405.

\bibitem{277132}
\BIBentryALTinterwordspacing
G.~Cherubin, R.~Jansen, and C.~Troncoso, ``Online website fingerprinting:
  Evaluating website fingerprinting attacks on tor in the real world,'' in
  \emph{31st USENIX Security Symposium (USENIX Security 22)}.\hskip 1em plus
  0.5em minus 0.4em\relax Boston, MA: USENIX Association, Aug. 2022, pp.
  753--770. [Online]. Available:
  \url{https://www.usenix.org/conference/usenixsecurity22/presentation/cherubin}
\BIBentrySTDinterwordspacing

\bibitem{puppeteer}
T.~G. Corporation, ``Tools for web developers: Puppeteer,''
  {https://developers.google.com/web/tools/puppeteer}.

\bibitem{dingledine2004tor}
R.~Dingledine, N.~Mathewson, P.~F. Syverson \emph{et~al.}, ``Tor: The
  second-generation onion router.'' in \emph{USENIX security symposium},
  vol.~4, 2004, pp. 303--320.

\bibitem{DraperGil2016CharacterizationOE}
\BIBentryALTinterwordspacing
G.~Draper-Gil, A.~H. Lashkari, M.~S.~I. Mamun, and A.~A. Ghorbani,
  ``Characterization of encrypted and vpn traffic using time-related
  features,'' in \emph{International Conference on Information Systems Security
  and Privacy}, 2016. [Online]. Available:
  \url{https://api.semanticscholar.org/CorpusID:21535780}
\BIBentrySTDinterwordspacing

\bibitem{197185}
\BIBentryALTinterwordspacing
J.~Hayes and G.~Danezis, ``k-fingerprinting: A robust scalable website
  fingerprinting technique,'' in \emph{25th USENIX Security Symposium (USENIX
  Security 16)}.\hskip 1em plus 0.5em minus 0.4em\relax Austin, TX: USENIX
  Association, Aug. 2016, pp. 1187--1203. [Online]. Available:
  \url{https://www.usenix.org/conference/usenixsecurity16/technical-sessions/presentation/hayes}
\BIBentrySTDinterwordspacing

\bibitem{he2005routing}
Y.~He, M.~Faloutsos, S.~Krishnamurthy, and B.~Huffaker, ``On routing asymmetry
  in the internet,'' in \emph{GLOBECOM'05. IEEE Global Telecommunications
  Conference, 2005.}, vol.~2.\hskip 1em plus 0.5em minus 0.4em\relax IEEE,
  2005, pp. 6--pp.

\bibitem{10.1145/1655008.1655013}
\BIBentryALTinterwordspacing
D.~Herrmann, R.~Wendolsky, and H.~Federrath, ``Website fingerprinting:
  attacking popular privacy enhancing technologies with the multinomial
  na\"{\i}ve-bayes classifier,'' in \emph{Proceedings of the 2009 ACM Workshop
  on Cloud Computing Security}, ser. CCSW '09.\hskip 1em plus 0.5em minus
  0.4em\relax New York, NY, USA: Association for Computing Machinery, 2009, p.
  31–42. [Online]. Available: \url{https://doi.org/10.1145/1655008.1655013}
\BIBentrySTDinterwordspacing

\bibitem{hestness2017deep}
J.~Hestness, S.~Narang, N.~Ardalani, G.~Diamos, H.~Jun, H.~Kianinejad, M.~M.~A.
  Patwary, Y.~Yang, and Y.~Zhou, ``Deep learning scaling is predictable,
  empirically,'' \emph{arXiv preprint arXiv:1712.00409}, 2017.

\bibitem{DBLP:conf/cikm/KanT05}
\BIBentryALTinterwordspacing
M.~Kan and H.~O.~N. Thi, ``Fast webpage classification using {URL} features,''
  in \emph{Proceedings of the 2005 {ACM} {CIKM} International Conference on
  Information and Knowledge Management, Bremen, Germany, October 31 - November
  5, 2005}, O.~Herzog, H.~Schek, N.~Fuhr, A.~Chowdhury, and W.~Teiken,
  Eds.\hskip 1em plus 0.5em minus 0.4em\relax {ACM}, 2005, pp. 325--326.
  [Online]. Available: \url{https://doi.org/10.1145/1099554.1099649}
\BIBentrySTDinterwordspacing

\bibitem{NIPS2012_c399862d}
\BIBentryALTinterwordspacing
A.~Krizhevsky, I.~Sutskever, and G.~E. Hinton, ``Imagenet classification with
  deep convolutional neural networks,'' in \emph{Advances in Neural Information
  Processing Systems}, F.~Pereira, C.~Burges, L.~Bottou, and K.~Weinberger,
  Eds., vol.~25.\hskip 1em plus 0.5em minus 0.4em\relax Curran Associates,
  Inc., 2012. [Online]. Available:
  \url{https://proceedings.neurips.cc/paper_files/paper/2012/file/c399862d3b9d6b76c8436e924a68c45b-Paper.pdf}
\BIBentrySTDinterwordspacing

\bibitem{Geekbench}
P.~labs, ``Geekbench browser,'' {https://www.geekbench.com/}.

\bibitem{li2022backdoor}
Y.~Li, Y.~Jiang, Z.~Li, and S.-T. Xia, ``Backdoor learning: A survey,'' 2022.

\bibitem{liberatore2006inferring}
M.~Liberatore and B.~N. Levine, ``Inferring the source of encrypted http
  connections,'' in \emph{Proceedings of the 13th ACM conference on Computer
  and communications security}, 2006, pp. 255--263.

\bibitem{lim_packet_2019}
H.-K. Lim, J.-B. Kim, J.-S. Heo, K.~Kim, Y.-G. Hong, and Y.-H. Han,
  ``Packet-based network traffic classification using deep learning,'' in
  \emph{2019 International Conference on Artificial Intelligence in Information
  and Communication (ICAIIC)}, 2019, pp. 046--051.

\bibitem{liu2023taxonomystructured}
T.~Liu, Z.~Xu, H.~He, G.-Y. Hao, G.-H. Lee, and H.~Wang, ``Taxonomy-structured
  domain adaptation,'' 2023.

\bibitem{lopez-2017-network}
M.~Lopez-Martin, B.~Carro, A.~Sanchez-Esguevillas, and J.~Lloret, ``Network
  traffic classifier with convolutional and recurrent neural networks for
  internet of things,'' \emph{IEEE access}, vol.~5, pp. 18\,042--18\,050, 2017.

\bibitem{lotfollahi2018deep}
M.~Lotfollahi, R.~S.~H. Zade, M.~J. Siavoshani, and M.~Saberian, ``Deep packet:
  A novel approach for encrypted traffic classification using deep learning,''
  2018.

\bibitem{similarweb}
S.~LTD, ``top website rank,'' {https://www.similarweb.com/}.

\bibitem{lu2010website}
L.~Lu, E.-C. Chang, and M.~C. Chan, ``Website fingerprinting and identification
  using ordered feature sequences,'' in \emph{Computer Security--ESORICS 2010:
  15th European Symposium on Research in Computer Security, Athens, Greece,
  September 20-22, 2010. Proceedings 15}.\hskip 1em plus 0.5em minus
  0.4em\relax Springer, 2010, pp. 199--214.

\bibitem{moore_network_traffic}
A.~K.~J. Michael, E.~Valla, N.~S. Neggatu, and A.~W. Moore, ``Network traffic
  classification via neural networks,''
  https://www.cl.cam.ac.uk/techreports/UCAM-CL-TR-912.pdf.

\bibitem{oshea2015introduction}
K.~O'Shea and R.~Nash, ``An introduction to convolutional neural networks,''
  2015.

\bibitem{Panchenko2016WebsiteFA}
\BIBentryALTinterwordspacing
A.~Panchenko, F.~Lanze, J.~Pennekamp, T.~Engel, A.~Zinnen, M.~Henze, and
  K.~Wehrle, ``Website fingerprinting at internet scale,'' in \emph{Network and
  Distributed System Security Symposium}, 2016. [Online]. Available:
  \url{https://api.semanticscholar.org/CorpusID:15302617}
\BIBentrySTDinterwordspacing

\bibitem{10.1145/3457904}
\BIBentryALTinterwordspacing
E.~Papadogiannaki and S.~Ioannidis, ``A survey on encrypted network traffic
  analysis applications, techniques, and countermeasures,'' \emph{ACM Comput.
  Surv.}, vol.~54, no.~6, jul 2021. [Online]. Available:
  \url{https://doi.org/10.1145/3457904}
\BIBentrySTDinterwordspacing

\bibitem{google_rezaei_2018achieve}
S.~Rezaei and X.~Liu, ``How to achieve high classification accuracy with just a
  few labels: A semi-supervised approach using sampled packets,'' \emph{arXiv
  preprint arXiv:1812.09761}, 2018.

\bibitem{Rezaei_2020}
\BIBentryALTinterwordspacing
------, ``Multitask learning for network traffic classification,'' in
  \emph{2020 29th International Conference on Computer Communications and
  Networks ({ICCCN})}.\hskip 1em plus 0.5em minus 0.4em\relax {IEEE}, aug 2020.
  [Online]. Available: \url{https://doi.org/10.1109%2Ficccn49398.2020.9209652}
\BIBentrySTDinterwordspacing

\bibitem{Rimmer_2018}
\BIBentryALTinterwordspacing
V.~Rimmer, D.~Preuveneers, M.~Juarez, T.~V. Goethem, and W.~Joosen, ``Automated
  website fingerprinting through deep learning,'' in \emph{Proceedings 2018
  Network and Distributed System Security Symposium}, ser. NDSS 2018.\hskip 1em
  plus 0.5em minus 0.4em\relax Internet Society, 2018. [Online]. Available:
  \url{http://dx.doi.org/10.14722/ndss.2018.23105}
\BIBentrySTDinterwordspacing

\bibitem{sahoo2019malicious}
D.~Sahoo, C.~Liu, and S.~C.~H. Hoi, ``Malicious url detection using machine
  learning: A survey,'' 2019.

\bibitem{7925139}
M.~Shafiq, X.~Yu, A.~A. Laghari, L.~Yao, N.~K. Karn, and F.~Abdessamia,
  ``Network traffic classification techniques and comparative analysis using
  machine learning algorithms,'' in \emph{2016 2nd IEEE International
  Conference on Computer and Communications (ICCC)}, 2016, pp. 2451--2455.

\bibitem{sirinam2018deep}
P.~Sirinam, M.~Imani, M.~Juarez, and M.~Wright, ``Deep fingerprinting:
  Undermining website fingerprinting defenses with deep learning,'' 2018.

\bibitem{sirinam2019triplet}
P.~Sirinam, N.~Mathews, M.~S. Rahman, and M.~Wright, ``Triplet fingerprinting:
  More practical and portable website fingerprinting with n-shot learning,'' in
  \emph{Proceedings of the 2019 ACM SIGSAC Conference on Computer and
  Communications Security}, 2019, pp. 1131--1148.

\bibitem{sun2017revisiting}
C.~Sun, A.~Shrivastava, S.~Singh, and A.~Gupta, ``Revisiting unreasonable
  effectiveness of data in deep learning era,'' in \emph{Proceedings of the
  IEEE international conference on computer vision}, 2017, pp. 843--852.

\bibitem{turner2019cleanlabel}
\BIBentryALTinterwordspacing
A.~Turner, D.~Tsipras, and A.~Madry, ``Clean-label backdoor attacks,'' 2019.
  [Online]. Available: \url{https://openreview.net/forum?id=HJg6e2CcK7}
\BIBentrySTDinterwordspacing

\bibitem{wang2020continuously}
H.~Wang, H.~He, and D.~Katabi, ``Continuously indexed domain adaptation,''
  2020.

\bibitem{wang2014effective}
T.~Wang, X.~Cai, R.~Nithyanand, R.~Johnson, and I.~Goldberg, ``Effective
  attacks and provable defenses for website fingerprinting,'' in \emph{23rd
  USENIX Security Symposium (USENIX Security 14)}, 2014, pp. 143--157.

\bibitem{10.1145/2517840.2517851}
\BIBentryALTinterwordspacing
T.~Wang and I.~Goldberg, ``Improved website fingerprinting on tor,'' in
  \emph{Proceedings of the 12th ACM Workshop on Workshop on Privacy in the
  Electronic Society}, ser. WPES '13.\hskip 1em plus 0.5em minus 0.4em\relax
  New York, NY, USA: Association for Computing Machinery, 2013, p. 201–212.
  [Online]. Available: \url{https://doi.org/10.1145/2517840.2517851}
\BIBentrySTDinterwordspacing

\bibitem{Wang_2022_CVPR}
Z.~Wang, J.~Zhai, and S.~Ma, ``Bppattack: Stealthy and efficient trojan attacks
  against deep neural networks via image quantization and contrastive
  adversarial learning,'' in \emph{Proceedings of the IEEE/CVF Conference on
  Computer Vision and Pattern Recognition (CVPR)}, June 2022, pp.
  15\,074--15\,084.

\bibitem{williams2006preliminary}
N.~Williams, S.~Zander, and G.~Armitage, ``A preliminary performance comparison
  of five machine learning algorithms for practical ip traffic flow
  classification,'' \emph{ACM SIGCOMM Computer Communication Review}, vol.~36,
  no.~5, pp. 5--16, 2006.

\bibitem{xu2023domainindexing}
Z.~Xu, G.-Y. Hao, H.~He, and H.~Wang, ``Domain-indexing variational bayes:
  Interpretable domain index for domain adaptation,'' 2023.

\bibitem{xu2023graphrelational}
Z.~Xu, H.~He, G.-H. Lee, Y.~Wang, and H.~Wang, ``Graph-relational domain
  adaptation,'' 2023.

\bibitem{zhuang2020comprehensive}
F.~Zhuang, Z.~Qi, K.~Duan, D.~Xi, Y.~Zhu, H.~Zhu, H.~Xiong, and Q.~He, ``A
  comprehensive survey on transfer learning,'' 2020.

\end{thebibliography}
